\documentclass[referee, useAMS, usenatbib, amstex, amssymb]{mnras}
\usepackage{psfig}
\usepackage{graphicx}
\usepackage{epsf}
\usepackage{bm}
\usepackage{lscape}
\usepackage[usenames, dvips]{color}
\title [Interrelated Main-Sequence Relations]
{Interrelated Main-Sequence Mass-Luminosity, Mass-Radius and Mass-Effective Temperature Relations}
\author[Eker~et~al.]
       {Z. Eker $^{1}$\thanks{E-mail: eker@akdeniz.edu.tr},
V. Bak{\i}\c{s}$^{1}$, S. Bilir$^{2}$, F. Soydugan$^{3,4}$, {I. Steer}$^{5}$,
E. Soydugan$^{3,4}$, 
\newauthor
H. Bak{\i}\c{s}$^{1}$, F. Ali\c{c}avu\c{s}$^{3,4}$, G. Aslan$^{1}$, M. Alpsoy$^{1}$
\\
    $^1$Akdeniz University, Faculty of Sciences, Department of Space Sciences and 
Technologies, 07058, Antalya, Turkey\\
    $^2$Istanbul University, Faculty of Science, Department of Astronomy and Space
Sciences, 34119, Istanbul, Turkey\\
  $^3$Department of Physics, Faculty of Arts and Sciences, \c{C}anakkale Onsekiz 
Mart University, 17100 \c{C}anakkale, Turkey\\
  $^4$Astrophysics Research Center and Ulup{\i}nar Observatory, \c{C}anakkale 
Onsekiz Mart University, 17100, \c{C}anakkale, Turkey\\
  $^5$NASA/IPAC Extragalactic Database of Distances, Toronto, Ontario, Canada\\
}
\date{}

\pagerange{\pageref{firstpage}--\pageref{lastpage}} \pubyear{}

\begin{document}

\maketitle

\label{firstpage} 
\begin{abstract}

Absolute parameters of 509 main-sequence stars selected from the components of detached-eclipsing spectroscopic binaries in the Solar neighbourhood are used to study mass-luminosity, mass-radius and mass-effective temperature relations (MLR, MRR and MTR). The MLR function is found better if expressed by a six-piece classical MLR ($L \propto M^{\alpha}$) rather than a fifth or a sixth degree polynomial within the mass range of $0.179\leq M/M_{\odot}\leq 31$. The break points separating the mass-ranges with classical MLR do not appear to us to be arbitrary. Instead, the data indicate abrupt changes along the mass axis in the mean energy generation per unit of stellar mass. Unlike the MLR function, the MRR and MTR functions cannot be determined over the full range of masses. A single piece MRR function is calibrated from the radii of stars with $M\leq1.5M_{\odot}$, while a second single piece MTR function is found for stars with $M>1.5M_{\odot}$. The missing part of the MRR is computed from the MLR and MTR, while the missing part of the MTR is computed from the MLR and MRR. As a result, we have interrelated MLR, MRR and MTR, which are useful in determining the typical absolute physical parameters of main-sequence stars of given masses. These functions are also useful to estimate typical absolute physical parameters from typical $T_{eff}$ values. Thus, we were able to estimate the typical absolute physical parameters of main-sequence stars observed in the Sejong Open Cluster survey, based on that survey's published values for $T_{eff}$. Since typical absolute physical parameters of main sequence stars cannot normally be determined in such photometric surveys, the interrelated functions are shown to be useful to compute such missing parameters from similar surveys.
\end{abstract}


\section{Introduction}
The main-sequence mass-luminosity relation (MLR), discovered by \citet{Hertzsprung23} and \citet*{Russell23} independently in the first half of the 20th century, is one of the fundamentally confirmed and universally recognized astronomical relations. Throughout the century it has been revised, updated and improved upon as newer and more accurate data became available. Among the many authors contributing to those revisions, some of the most notable include \citet{Eddington26}, \citet{McLaughlin27}, \citet{Kuiper38}, \citet{Petrie50a, Petrie50b}, \citet{Strand54}, \citet{Eggen56}, \citet{Popper67, Popper80}, \citet{McCluskey72}, \citet{Heintz78}, \citet*{Cester83}, \citet*{Griffiths88}, \citet{Henry93}, \citet{Demircan91}, \citet{Andersen91}, \citet{Gorda98}, \citet{Ibanoglu06}, \citet{Malkov07}, \citet*{Torres10} and \citet*{Gafeira12}. Recently, \citet{Eker15}, \citet{Benedict16} and \citet{Moya18} have also contributed to those revisions. Empirical MLRs are useful to provide an easy and independent way of testing the absolute brightness or parallax of a main-sequence star once its mass is estimated. Conversely, a MLR is also useful, to estimate masses of single main-sequence stars from their observationally determined luminosities. Furthermore, an accurate MLR has practical applications in extragalactic research, since it can be used along with the stellar content of a galaxy, to estimate the stellar mass of a galaxy.

In contrast, mass-radius relations (MRR) for main-sequence stars began to appear in the literature only after the mid-20th century \citep{McCrea50, Plaut53, Huang56, Lacy77, Lacy79, Kopal78, Patterson84, Gimenez85, Harmanec88, Demircan91, Malkov07}. Note further that the Stefan-Boltzmann law clearly indicates stellar luminosities are related to stellar radii and effective temperatures. Having empirically determined MLR and MRR available, one is not free to determine another independent mass-effective temperature relation (MTR). Thus, independent MTR determinations have not yet been attempted. 

Unfortunately, there is no consensus on what functions properly express the MLR and the MRR. Most early research studied relations between bolometric absolute magnitude ($M_{bol}$) and mass ($\log M$), while later authors studied relations between $\log L$ and $\log M$. The most common relations are linear \citep{Eggen56, McCluskey72, Heintz78, Kopal78, Cester83, Griffiths88, Demircan91, Henry93} and quadratic \citep{Petrie50a, Petrie50b, Demircan91, Henry93, Fang10}. In addition, there have been studies into the use of third or fourth degree polynomials for the MLR or MRR \citep{Demircan91, Malkov07, Gafeira12}. 

Almost a century-long history and the latest developments regarding the MLR and MRR functions have been summarized and revised by \citet{Eker15}. MLR revisions are based on the simplest form ($L \propto M^{\alpha}$) for a subset of Galactic nearby main-sequence stars with masses and radii accurate to $\leq3\%$ and luminosities accurate to $\leq30\%$. The basic stellar astrophysical data ($M$, $R$ and $T_{eff}$) were taken from ``The Catalog of Stellar Parameters from the Detached Double-Lined Eclipsing Binaries in the Milky Way'' by \citet{Eker14}. The revised MLRs were determined according to a constant power law ($\propto M^{\alpha}$) within four distinct stellar mass domains (low mass: $0.38\leq M/M_{\odot}\leq1.05$, intermediate mass: $1.05<M/M_{\odot}\leq2.40$, high mass: $2.4<M/M_{\odot}\leq7$, and very high mass: $7<M/M_{\odot}\leq32$), identified according to the efficiency of stellar energy production per stellar mass ($L/M$). The four-piece linear MLRs were argued to be preferable to a single linear, quadratic or cubic relation within the total range of stellar masses studied $0.38<M/M_{\odot}\leq32$. 

The mass-radius and the mass-temperature diagrams were studied by \citet{Eker15}, but derivations of MRR or MTR were not attempted. While stars with masses of $M<1M_{\odot}$ exhibit a very narrow distribution of radii, stars with masses of $M>1M_{\odot}$ in contrast, exhibit a very broad distribution of radii. No single function therefore, was found suitable to express the MRR of main-sequence stars in the M-R diagram. Like the earlier studies, which were satisfied with MLR only, calibration of MTR was not considered. 

Crucially however, the MRR and MTR functions, as well as the MLR functions, are needed by the astronomical community for practical purposes. Those include the need to be able to estimate a typical luminosity, radius and $T_{eff}$ for main-sequence stars of a given mass. Despite such a common need, however, a database of typical luminosities, temperatures, radii and spectral types of the main-sequence stars in the Solar neighbourhood has not yet been fully compiled. Indeed, most of the astronomical community still commonly uses the two tables for main-sequence stars by \citet{Cox00}, which were compiled from \citet{Hoxie73, Lacy77, Schmidt82, Johnson66} and \citet{dejager87}. One of the tables \citep[Table 15.7 in][]{Cox00} gives calibration of absolute magnitudes of MK spectral types (Spectral type, $M_V$, $B-V$, $R-I$, $T_{eff}$ and $BC$). The other table \citep[Table 15.8 in][]{Cox00}, however, indicates typical mass ($M$), radius ($R$), $\log g$, mean density and rotation speed for a main-sequence star of a given spectral type with a footnote: ``A column indicates an uncertain value''. Obviously, the column with spectral types was chosen as a tie or a connecting column between the two tables where the former one is sufficiently reliable but the later one is not. Therefore, calibration tables with spectral types, $T_{eff}$, colours and bolometric corrections, such as a table produced by \citet{Sung13}, could not be associated with absolute parameters such as masses ($M$) and radii ($R$) with sufficient accuracy.  

In this study, the empirical MLRs for main-sequence stars by \citet{Eker15} are updated, and extended in order to simultaneously determine new MRR and MTR functions. The derived functions (MLR, MRR and MTR), we claim, do indeed produce a typical luminosity, a typical radius, and a typical effective temperature that is consistent with the Stefan-Boltzmann law ($L=4\pi R^2 \sigma T^4_{eff}$) for a typical main-sequence star of a given mass. That is, newly determined MLR, MRR and MTR functions are not independent but interrelated, so they can be used for assigning an absolute $M$ and $R$ with sufficient accuracy to the spectral types and $T_{eff}$ given in observationally determined calibration tables, such as the one given by \citet{Sung13}, for the Solar neighbourhood main-sequence stars.

\section{The Data}
\subsection{The preliminary sample}
The primary data source of this study is the updated ``Catalog of Stellar Parameters from the Detached Double-Lined Eclipsing Binaries in the Milky Way'' by \citet{Eker14}. The older version, which provided parameters published up to January 2, 2013, was updated to include parameters published up to January 2, 2017. The number of detached double lined eclipsing binaries increased from 257 to 319, even after removing the three systems (GZ Leo, DH Cep, NSVS 01031772), since they were found not fitting to the criteria of \citet{Eker14}. In all therefore, 65 new systems were included, while masses ($M$), radii ($R$) and effective temperatures ($T_{eff}$) for 33 systems in the older version were renewed with the new values published after January 2013 and up to January 2017. 

The total number of stars is now 639, when the components of the binaries are counted separately. The number is not even because one of the new systems (TYC 6212-1273-1) is an eclipsing detached spectroscopic triple (SB3).  With 125 new stars added to the old version, the number of stars is increased by 24\%. Further, in addition to increasing the quantity of our data, there has also been an increase  in the quality. The number of stars with both $M$ and $R$ measurements better than 1\% uncertainty is increased from 93 to 134. The number of stars with both $M$ and $R$ measurements better than 3\% uncertainty is increased from 311 to 400. The number of stars with both $M$ and $R$ measurements better than 5\% uncertainty is increased from 388 to 480. 

Unfortunately, not all stars in the catalog have published $T_{eff}$ because some authors \citep{Young06, Shkolnik08, Helminiak09, Sandquist13} prefer to give temperature ratios rather than component temperatures as solutions of observed radial velocity and light curves of the detached eclipsing binaries. Therefore 586 stars with published $T_{eff}$ in the updated catalogue were chosen as the preliminary sample for this study. 

The basic astrophysical parameters of 586 stars in the preliminary sample are listed in Table 1. The columns are self explanatory to indicate: identification number (ID), name, component ID, celestial coordinates (International Celestial Reference System in J2000.0), spectral type, reference; $M$, error of $M$; $R$, error of $R$; $\log g$, error of $\log g$; reference; $T_{eff}$, error of $T_{eff}$; reference and remarks. Usually, the spectral types, $M$, $R$, $\log g$, and $T_{eff}$ of the components of a detached binary are found in a single reference. However, rarely, some velocity and light curve solutions appear without temperatures and/or spectral types \citep{Young06, Shkolnik08, Helminiak09, Sandquist13} giving temperature ratios rather than individual temperatures. Thus we had to assign three columns in Table 1 as references for the columns before them. Moreover, $M$ and $R$ collected from older references are homogenized and re-evaluated using recently updated and more accurate constants $GM_{\odot}=1.3271244\times10^{20}$ m$^3$s$^{-2}$ \citep{Standish95} and $R_{\odot}=6.9566\times10^8$m \citep*{Haberreiter08} by \citet{Eker14}. Therefore, absolute parameters $M$, $R$ and $\log g$ coming from the older references are given a single reference. Interested readers may follow the references given in \citet{Eker14} for the original published values.  

\begin{landscape}
\begin{table}
{\scriptsize
\setlength{\tabcolsep}{2pt}
\centering
\caption{Most accurate masses and radii with published $T_{eff}$ from the detached double-lined eclipsing binaries in the Solar neighbourhood and within the Galactic disc.}
\begin{tabular}{rccccccccccccccccc}
\hline
ID & Star name & Comp. & $\alpha$ (J2000) & $\delta$(J2000) & Spt Type & Reference & $M$ & $M_{err}$ & $R$ & $R_{err}$ & $\log g$ & $\log g_{err}$ & Reference  & $T_{eff}$ & $T_{err}$ & Reference & Remarks$^{(a)}$ \\
  &  &  & (hh:mm:mm.ss) & (dd:mm:ss.ss) &  & & ($M_{\odot}$) & ($M_{\odot}$) & ($R_{\odot}$) & ($R_{\odot}$) & (cgs) & (cgs) &   & (K) & (K) &  & \\
\hline

 1  & V421 Peg & p & 00:07:02.00 & +22:50:40.03 & F1V   & 2016NewA...46...47O     & 1.594 & 0.029 & 1.584 & 0.028 & 4.241 & 0.017 & 2016NewA...46...47O & 7250  & 80    & 2016NewA...46...47O & * \\
 2  & V421 Peg & s & 00:07:02.00 & +22:50:40.03 & F2V   & 2016NewA...46...47O     & 1.356 & 0.029 & 1.328 & 0.029 & 4.324 & 0.021 & 2016NewA...46...47O & 6980  & 120   & 2016NewA...46...47O & * \\
 3  & V342 And B & p & 00:10:03.68 & +46:23:25.80 & $--$& $--$                    & 1.270 & 0.010 & 1.210 & 0.010 & 4.377 & 0.008 & 2015A\&A...575A.101D& 6395  & $--$  & 2015A\&A...575A.101D & * \\
 4  & V342 And B & s & 00:10:03.68 & +46:23:25.80 & $--$& $--$                    & 1.280 & 0.010 & 1.250 & 0.010 & 4.352 & 0.008 & 2015A\&A...575A.101D& 6200  & 30    & 2015A\&A...575A.101D & * \\
 5  & DV Psc & p & 00:13:09.20 & +05:35:43.06 & K4V   & 2007MNRAS.382.1133Z       & 0.677 & 0.019 & 0.685 & 0.030 & 4.598 & 0.040 & 2014PASA...31...24E & 4450  & 8     & 2007MNRAS.382.1133Z & ** \\
 6  & DV Psc & s & 00:13:09.20 & +05:35:43.06 & M1V   & 2007MNRAS.382.1133Z       & 0.475 & 0.010 & 0.514 & 0.020 & 4.693 & 0.035 & 2014PASA...31...24E & 3614  & 8     & 2007MNRAS.382.1133Z & ** \\
 7  & MU Cas & p & 00:15:51.56 & +60:25:53.64 & B5V   & 2004AJ....128.1840L       & 4.657 & 0.100 & 4.192 & 0.050 & 3.862 & 0.014 & 2014PASA...31...24E & 14750 & 500   & 2004AJ....128.1840L & * \\
 8  & MU Cas & s & 00:15:51.56 & +60:25:53.64 & B5V   & 2004AJ....128.1840L       & 4.575 & 0.090 & 3.671 & 0.040 & 3.969 & 0.013 & 2014PASA...31...24E & 15100 & 500   & 2004AJ....128.1840L & * \\
 9  & GSC 4019 3345 & p & 00:22:45.37 & +62:20:05.50 & A4V   & 2013PASA...30...26B & 1.920 & 0.010 & 1.760 & 0.050 & 4.231 & 0.025 & 2013PASA...30...26B & 8600  & 310   & 2013PASA...30...26B & * \\
10  & GSC 4019 3345 & s & 00:22:45.37 & +62:20:05.50 & A4V   & 2013PASA...30...26B & 1.920 & 0.010 & 1.760 & 0.050 & 4.231 & 0.025 & 2013PASA...30...26B & 8600  & 570   & 2013PASA...30...26B & * \\
... & ...      & ...   & ... & ...  & ...  & ... & ... & ... & ... & ... & ... & ... & ... & ... & ... & ... & ... \\
... & ...      & ...   & ... & ...  & ...  & ... & ... & ... & ... & ... & ... & ... & ... & ... & ... & ... & ... \\
... & ...      & ...   & ... & ...  & ...  & ... & ... & ... & ... & ... & ... & ... & ... & ... & ... & ... & ... \\
577 & IT Cas   & p & 23:42:01.40 & +51:44:36.80 & F6V   & 1997AJ....114.1206L  & 1.330 & 0.009 & 1.603 & 0.015 & 4.152 & 0.009 & 2014PASA...31...24E & 6470  & 110   & 1997AJ....114.1206L & * \\
578 & IT Cas   & s & 23:42:01.40 & +51:44:36.80 & F6V   & 1997AJ....114.1206L  & 1.328 & 0.008 & 1.569 & 0.040 & 4.170 & 0.022 & 2014PASA...31...24E & 6470  & 110   & 1997AJ....114.1206L & * \\
579 & BK Peg   & p & 23:47:08.46 & +26:33:59.92 & F8    & 1983AJ.....88.1242P  & 1.414 & 0.007 & 1.985 & 0.008 & 3.993 & 0.004 & 2014PASA...31...24E & 6265  & 85    & 2010A\&A...516A..42C & * \\
580 & BK Peg   & s & 23:47:08.46 & +26:33:59.92 & F8    & 1983AJ.....88.1242P  & 1.257 & 0.005 & 1.472 & 0.017 & 4.202 & 0.010 & 2014PASA...31...24E & 6320  & 30    & 2010A\&A...516A..42C & * \\
581 & AP And   & p & 23:49:30.71 & +45:47:21.25 & F6    & 1984ApJS...54..421G  & 1.277 & 0.004 & 1.234 & 0.006 & 4.362 & 0.005 & 2014AJ....147..148L & 6565  & 150   & 2014AJ....147..148L & * \\
582 & AP And   & s & 23:49:30.71 & +45:47:21.25 & F8    & 1984ApJS...54..421G  & 1.251 & 0.004 & 1.196 & 0.005 & 4.381 & 0.004 & 2014AJ....147..148L & 6495  & 150   & 2014AJ....147..148L & * \\
583 & AL Scl   & p & 23:55:16.58 & -31:55:17.28 & B6V   & 1987A\&A...179..141H & 3.617 & 0.110 & 3.241 & 0.050 & 3.975 & 0.019 & 2014PASA...31...24E & 13550 & 350   & 1987A\&A...179..141H & ** \\
584 & AL Scl   & s & 23:55:16.58 & -31:55:17.28 & B9V   & 1987A\&A...179..141H & 1.703 & 0.040 & 1.401 & 0.020 & 4.377 & 0.016 & 2014PASA...31...24E & 10300 & 360   & 1987A\&A...179..141H & * \\
585 & V821 Cas & p & 23:58:49.17 & +53:40:19.82 & A1.5V & 2009MNRAS.395.1649C  & 2.025 & 0.066 & 2.308 & 0.028 & 4.018 & 0.018 & 2014PASA...31...24E & 9400  & 400   & 2009MNRAS.395.1649C & ** \\
586 & V821 Cas & s & 23:58:49.17 & +53:40:19.82 & A4Vm  & 2009MNRAS.395.1649C  & 1.620 & 0.058 & 1.390 & 0.022 & 4.362 & 0.021 & 2014PASA...31...24E & 8600  & 400   & 2009MNRAS.395.1649C & ** \\
\hline
\end{tabular}%
\label{tab:addlabel}%
}
(a) $*$: very accurate ($M$ and $R$ errors $\leq 3\%$) \\
$**$: accurate ($M$ and $R$ errors 3-6\%) \\
$***$: less accurate ($M$ and $R$ errors 6-15\%) \\
1 : Discarded because $M$ or $R$ error(s) $>15\%$ \\
2 : Discarded according to the position on $M-R$ diagram (outside the limit defined by PARSEC models \citep{Bressan12} with $0.008\leq Z\leq0.004$) \\
3 : Discarded since oversized and hotter than normal main-sequence stars \citep{Iglesias-Marzoa17} \\
GLB : Discarded because of member of a globular cluster \\
\end{table}%
\end{landscape}

\subsection{The sample and its constraints}

The preliminary sample of 586 stars is not homogeneous, as required for reliable study of the MLR, MRR, and MTR functions. In fact, it is heterogeneous in at least one respect. It contains only mostly Solar neighbourhood disc stars, as well as some globular cluster stars. Indeed, the preliminary sample is heterogeneous in another respect. It contains only mostly main-sequence stars, as well as some non-main-sequence stars with relative $M$ errors showing a peak at 1\% and relative $R$ errors showing a peak at 2\%, which are similar to the error distributions studied by \citet{Eker14}. Therefore, in order to study the MLR, MRR and MTR functions of purely main-sequence stars located specifically in the Solar neighbourhood and within the Galactic disc, additional constraints in our sample selection procedures are required. The first constraint to apply is the relative errors of $M$ and $R$.

\subsubsection{Constraining relative errors of $M$ and $R$}

Limiting accuracy on observational $M$ and $R$ values are important for astrophysical point of view in order to compare stellar structure and evolution models with actual observations. Therefore it is common to have a limiting accuracy decided by researchers according to the nature of the study involved, nevertheless it is arbitrary in most cases. \citet{Andersen91} preferred to collect detached, double lined binary systems with $M$ and $R$ measurements accurate to $\leq 2\%$. \citet{Torres10} slightly extended this limit to $\leq3\%$. However, studying MLR of low mass stars \citet{Henry93} allowed observational masses accurate up to 15\%. On the other hand, studying MLR and MRR in the ranges $0.63\leq M/M_{\odot}\leq 31.6$, $0.63\leq R/R_{\odot}\leq 25.1$, \citet{Malkov07} were satisfied with $M$ and $R$ accuracies up to 10\%. ``Single-parameter relations used to predict $M$ and $R$ for single stars are limited to an accuracy of some $\pm 15\%$ in $M$ and $\pm 50\%$ in $R$, basically independent of the number and accuracy of the data used to establish the relations'' commented \citet{Andersen91}. Therefore, for this study, we preferred the limiting accuracy to be 15\%. Since there are only 17 among 586 stars have $M$ or $R$ accuracies worse than 15\%, the number of stars in the study sample is reduced to 569 after the first constraint.      

\subsubsection{Constraining with respect to spatial distribution}

Constraining the study sample according to Galactic locations is also necessary because it has been known since almost a century now that the stars in the Galactic disc are mostly metal-rich Population I stars while the stars in globular clusters or in halo are metal-poor Population II stars, which show different mass-luminosity relations because of their metallicity and age differences. Searching through the preliminary sample, we have found 12 stars (six binaries) belonging to these three globular clusters: 47 Tuc \citep[one binary by ][]{Thompson10}, NGC 6362 \citep[two binaries by][]{Kaluzny15} and M4 \citep[three binaries by][]{Kaluzny13}. After removing these globular cluster members, the number of the stars in the study sample is reduced to 557 after the second constraint. 

\subsubsection{Constraining metallicity and age}

In order to determine MLR, MRR and MTR for main-sequence stars at Solar vicinity in the Galactic disc, we must also constrain our sample by identifying main-sequence stars within the observed distribution of metallicities. Unfortunately, the number of detached binaries with reliable metallicity is very limited. Among 176 detached binaries in the online database DEBCat\footnote{http://www.astro.keele.ac.uk/jkt/debcat/} by \citet{Southworth15}, metallicity information exists for only 66 systems, some of which are not real measurements but assumptions, and some of which have low accuracy. Studying the metallicity and age contributions to MLR for main-sequence stars, \citet*{Gafeira12} were able to work with only 13 binaries out of 94 in the list of \citet{Torres10}. Searching through our updated catalogue, we have found 72 stars (36 binaries) with spectroscopically determined [Fe/H] measurements which are listed in Table 2. Photometric metallicity determinations usually involve a number of uncertainties, including interstellar reddening. Therefore, we chose to include only spectroscopic determinations here. In most cases, they refer to the binary itself, but we have included a few systems in open clusters with [Fe/H] determinations from other cluster members. The number of [Fe/H] determinations is definitely not sufficient, because 72 stars with metallicities are not enough to calibrate a reliable MLR, MRR or MTR. However, these stars are still useful to indicate the metallicity distribution of the study sample.

\begin{landscape}
\begin{table}
\setlength{\tabcolsep}{2pt}
\centering
\caption{Spectroscopically determined relative iron abundances ([Fe/H] dex) of 72 stars, which are the components of detached eclipsing spectroscopic binaries in the Solar neighbourhood and within the Galactic disc.}
\begin{tabular}{cclcccr|cclcccr}
\hline
   &        &        &           &        &     &          &    &        &        &           &        &     &          \\
ID & Cat No & System & Comp. & [Fe/H] & $Z$ & Bib Code & ID & Cat No & System & Component & [Fe/H] & $Z$ & Bib Code \\
\hline
 1 &   1 & V421 Peg      & p & -0.11$\pm$0.08 & 0.0119$\pm$0.0021 & 2016NewA...46...47O  & 37 & 267 & ZZ Boo          & p & -0.10$\pm$0.08 & 0.0122$\pm$0.0022 & 2012AJ....144...35K  \\
 2 &   2 & V421 Peg      & s & -0.11$\pm$0.09 & 0.0119$\pm$0.0024 & 2016NewA...46...47O  & 38 & 268 & ZZ Boo          & s & -0.03$\pm$0.10 & 0.0142$\pm$0.0032 & 2012AJ....144...35K  \\
 3 &   3 & V342 And B    & p & -0.10$\pm$0.06 & 0.0122$\pm$0.0016 & 2015A\&A...575A.101D & 39 & 273 & V636 Cen        & p & -0.20$\pm$0.08 & 0.0098$\pm$0.0018 & 2009A\&A...502..253C \\
 4 &   4 & V342 And B    & s & -0.10$\pm$0.06 & 0.0122$\pm$0.0016 & 2015A\&A...575A.101D & 40 & 274 & V636 Cen        & s & -0.20$\pm$0.08 & 0.0098$\pm$0.0018 & 2009A\&A...502..253C \\
 5 &  13 & YZ Cas        & p &  0.54$\pm$0.11 & 0.0463$\pm$0.0097 & 2014MNRAS.438..590P  & 41 & 279 & AD Boo          & p &  0.10$\pm$0.15 & 0.0189$\pm$0.0062 & 2008A\&A...487.1095C \\
 6 &  14 & YZ Cas        & s &  0.01$\pm$0.11 & 0.0155$\pm$0.0038 & 2014MNRAS.438..590P  & 42 & 280 & AD Boo          & p &  0.10$\pm$0.15 & 0.0189$\pm$0.0062 & 2008A\&A...487.1095C \\
 7 &  15 & NGC188 KR V12 & p & -0.14          & 0.0112            & 2009AJ....137.5086M  & 43 & 323 & WZ Oph          & p & -0.27$\pm$0.07 & 0.0084$\pm$0.0013 & 2008A\&A...487.1095C \\
 8 &  16 & NGC188 KR V12 & s & -0.13          & 0.0114            & 2009AJ....137.5086M  & 44 & 324 & WZ Oph          & s & -0.27$\pm$0.07 & 0.0084$\pm$0.0013 & 2008A\&A...487.1095C \\
 9 &  41 & V505 Per      & p & -0.15          & 0.0109            & 2013PASP..125..753B  & 45 & 347 & V2653 Oph       & p & -0.11$\pm$0.08 & 0.0119$\pm$0.0021 & 2016NewA...45...36C  \\
10 &  42 & V505 Per      & s & -0.15          & 0.0109            & 2013PASP..125..753B  & 46 & 348 & V2653 Oph       & s &  0.07$\pm$0.09 & 0.0177$\pm$0.0034 & 2016NewA...45...36C  \\
11 &  47 & XY Cet        & p &  0.50          & 0.0429            & 2011MNRAS.414.3740S  & 47 & 375 & HD 172189       & p & -0.28          & 0.0082            & 2009A\&A...507..901C \\
12 &  48 & XY Cet        & s &  0.50          & 0.0429            & 2011MNRAS.414.3740S  & 48 & 376 & HD 172189       & s & -0.28          & 0.0082            & 2009A\&A...507..901C \\
13 &  57 & V570 Per      & p &  0.02$\pm$0.02 & 0.0159$\pm$0.0007 & 2010A\&A...516A..42C & 49 & 381 & CoRoT 105906206 & p &  0.00$\pm$0.10 & 0.0152$\pm$0.0034 & 2014A\&A...565A..55D \\
14 &  58 & V570 Per      & s &  0.02$\pm$0.02 & 0.0159$\pm$0.0007 & 2010A\&A...516A..42C & 50 & 382 & CoRoT 105906206 & s &  0.00$\pm$0.10 & 0.0152$\pm$0.0034 & 2014A\&A...565A..55D \\
15 &  79 & V1130 Tau     & p & -0.25$\pm$0.10 & 0.0088$\pm$0.0020 & 2010A\&A...510A..91C & 51 & 421 & V565 Lyr V18    & p &  0.31$\pm$0.06 & 0.0293$\pm$0.0036 & 2011A\&A...525A...2B \\
16 &  80 & V1130 Tau     & s & -0.25$\pm$0.10 & 0.0088$\pm$0.0020 & 2010A\&A...510A..91C & 52 & 422 & V565 Lyr V18    & s &  0.22$\pm$0.10 & 0.0243$\pm$0.0051 & 2011A\&A...525A...2B \\
17 & 109 & CD Tau        & p &  0.08$\pm$0.15 & 0.0181$\pm$0.0059 & 1999MNRAS.309..199R  & 53 & 423 & V568 Lyr V20    & p &  0.26$\pm$0.06 & 0.0264$\pm$0.0033 & 2011A\&A...525A...2B \\
18 & 110 & CD Tau        & s &  0.08$\pm$0.15 & 0.0181$\pm$0.0059 & 1999MNRAS.309..199R  & 54 & 424 & V568 Lyr V20    & s &  0.26$\pm$0.06 & 0.0264$\pm$0.0033 & 2011A\&A...525A...2B \\
19 & 141 & IM Mon        & p &  0.20$\pm$0.15 & 0.0233$\pm$0.0075 & 2011PASJ...63.1079B  & 55 & 447 & KIC 9777062     & p &  0.46$\pm$0.13 & 0.0396$\pm$0.0102 & 2016ApJ...831...11S  \\
20 & 142 & IM Mon        & s &  0.20$\pm$0.15 & 0.0233$\pm$0.0075 & 2011PASJ...63.1079B  & 56 & 448 & KIC 9777062     & s & -0.03$\pm$0.13 & 0.0142$\pm$0.0041 & 2016ApJ...831...11S  \\
21 & 143 & RR Lyn        & p &  0.31$\pm$0.08 & 0.0293$\pm$0.0049 & 2001ARep...45..888K  & 57 & 461 & HD 187669       & p & -0.25$\pm$0.10 & 0.0088$\pm$0.0020 & 2015MNRAS.448.1945H  \\
22 & 144 & RR Lyn        & s & -0.24$\pm$0.06 & 0.0090$\pm$0.0012 & 2001ARep...45..888K  & 58 & 462 & HD 187669       & s & -0.25$\pm$0.10 & 0.0088$\pm$0.0020 & 2015MNRAS.448.1945H  \\
23 & 151 & V501 Mon      & p &  0.01$\pm$0.06 & 0.0155$\pm$0.0021 & 2015AJ....150..154T  & 59 & 467 & KIC 9851944     & p & -0.06$\pm$0.05 & 0.0133$\pm$0.0015 & 2016ApJ...826...69G  \\
24 & 152 & V501 Mon      & s &  0.01$\pm$0.06 & 0.0155$\pm$0.0021 & 2015AJ....150..154T  & 60 & 468 & KIC 9851944     & s & -0.04$\pm$0.05 & 0.0139$\pm$0.0015 & 2016ApJ...826...69G  \\
25 & 155 & GX Gem        & p & -0.12$\pm$0.10 & 0.0117$\pm$0.0026 & 2008AJ....135.1757L  & 61 & 525 & OO Peg          & p & -0.10$\pm$0.01 & 0.0122$\pm$0.0003 & 2015NewA...37...70C  \\
26 & 156 & GX Gem        & s & -0.12$\pm$0.10 & 0.0117$\pm$0.0026 & 2008AJ....135.1757L  & 62 & 526 & OO Peg          & s &  0.05$\pm$0.02 & 0.0169$\pm$0.0008 & 2015NewA...37...70C  \\
27 & 165 & SW CMa        & p &  0.49$\pm$0.15 & 0.0420$\pm$0.0123 & 2012A\&A...537A.117T & 63 & 527 & NGC 7142 V2     & p & -0.03$\pm$0.06 & 0.0142$\pm$0.0019 & 2013AJ....146...40S  \\
28 & 166 & SW CMa        & s &  0.61$\pm$0.15 & 0.0528$\pm$0.0147 & 2012A\&A...537A.117T & 64 & 528 & NGC 7142 V2     & s & -0.12$\pm$0.02 & 0.0117$\pm$0.0005 & 2013AJ....146...40S  \\
29 & 167 & HW CMa        & p &  0.33$\pm$0.10 & 0.0305$\pm$0.0063 & 2012A\&A...537A.117T & 65 & 529 & V375 Cep        & p &  0.09$\pm$0.02 & 0.0185$\pm$0.0008 & 2013AJ....146...40S  \\
30 & 168 & HW CMa        & s &  0.28$\pm$0.10 & 0.0276$\pm$0.0057 & 2012A\&A...537A.117T & 66 & 530 & V375 Cep        & s &  0.09$\pm$0.02 & 0.0185$\pm$0.0008 & 2013AJ....146...40S  \\
31 & 199 & VZ Hya        & p & -0.20$\pm$0.12 & 0.0098$\pm$0.0027 & 2008A\&A...487.1095C & 67 & 541 & BW Aqr          & p & -0.07$\pm$0.11 & 0.0130$\pm$0.0032 & 2010A\&A...516A..42C \\
32 & 200 & VZ Hya        & s & -0.20$\pm$0.12 & 0.0098$\pm$0.0027 & 2008A\&A...487.1095C & 68 & 542 & BW Aqr          & s & -0.07$\pm$0.11 & 0.0130$\pm$0.0032 & 2010A\&A...516A..42C \\
33 & 203 & RS Cha        & p &  0.17$\pm$0.01 & 0.0219$\pm$0.0005 & 2005A\&A...442..993A & 69 & 557 & EF Aqr          & p &  0.00$\pm$0.10 & 0.0152$\pm$0.0034 & 2012A\&A...540A..64V \\
34 & 204 & RS Cha        & s &  0.17$\pm$0.01 & 0.0219$\pm$0.0005 & 2005A\&A...442..993A & 70 & 558 & EF Aqr          & s &  0.00$\pm$0.10 & 0.0152$\pm$0.0034 & 2012A\&A...540A..64V \\
35 & 253 & VV Crv        & p &  0.03          & 0.0162            & 2013AJ....146..146F  & 71 & 579 & BK Peg          & p & -0.12$\pm$0.07 & 0.0117$\pm$0.0018 & 2010A\&A...516A..42C \\
36 & 254 & VV Crv        & s &  0.03          & 0.0162            & 2013AJ....146..146F  & 72 & 580 & BK Peg          & s & -0.12$\pm$0.07 & 0.0117$\pm$0.0018 & 2010A\&A...516A..42C \\
\hline
\end{tabular}%
\label{tab:addlabel}%
\end{table}%
\end{landscape}

Figure 1 displays [Fe/H] distribution of Galactic disc stars which are candidates for the study sample. The mean, the median and the standard deviation of the distribution are +0.071, -0.015, and 0.218 dex respectively. Relying on statistics implied by this sub-sample (12.9\%), we may assume that most of the study sample have relative iron abundance [Fe/H]=0 dex distributed between the limits $-0.28\leq {\rm [Fe/H]}\leq 0.61$ dex. Thus, if the relative iron abundance of a star in the study sample is not known, one may assume it is most likely an approximately Solar one.
 
Relative iron abundance [Fe/H] is a good parameter to indicate metallicity of a star, however, it is not directly used by stellar structure and evolution models, which prefers to use parameter ($Z$) rather than [Fe/H], where ($Z$) represents the number ratio (percentage) of all elements other than hydrogen ($X$) and helium ($Y$), so $X+Y+Z=1$. Therefore, corresponding $Z$ values of [Fe/H] measurements were calculated and errors were propagated using the formulae given by Jo Bovy\footnote{https://github.com/jobovy/isodist/blob/master/isodist/Isochrone.py}, which is commonly used in transforming relative iron abundances of numerous open clusters and specifically used for the PARSEC models \citep{Bressan12}.

Stellar structure and evolution models predict luminosity ($L$) and radius ($R$) as directly observable parameters for a given stellar mass ($M$) and chemical composition ($X$, $Y$, $Z$) as a function of time (stellar tracks). $T_{eff}$ is not a parameter directly predicted. Instead, $T_{eff}$ given by models is computed from the predicted $L$ and $R$ according to the Stefan-Boltzmann law. Today, there are many theoretical stellar evolution models available. Notable examples include: Geneva Grids of Stellar Evolution Models \citep{Schaller92, Schaerer93},
Padova Database of Stellar Evolution Tracks \citep{Girardi00, Marigo08, Bertelli08, Bertelli09, Girardi10, Bressan12, Chen15}, Yonsei-Yale Isochrones \citep{Demarque04, Demarque08}, Victoria-Regina \citep*{Vandenberg06}, Dartmounth Stellar Evolution Database \citep{Dotter08}, the Pisa Stellar Evolution Database \citep*{Tognelli11}, Geneva Grids of Stellar Models with Rotation \citep{Meynet00, Ekstrom12, Mowlavi12, Georgy13}, Basti Stellar Evolution Database \citep {Pietrinferni13}, Mesa Isochrones and Stellar Tracks \citep{Paxton11, Paxton13, Paxton15, Dotter16, Choi16}, others such as \citet{Pols98}, \citet{Yildiz15}, and more.

PARSEC is an extended and updated version of the stellar evolution model previously used by \citet{Bressan81}, \citet{Girardi00}, and \citet{Bertelli08, Bertelli09}, as thoroughly described by \citet{Bressan12}. In order to differentiate peculiar stars with metal content outside the limits indicated by the histogram distribution (Fig. 1) and probable non-main sequence stars (pre or post main-sequence), we have also chosen to use zero age main-sequence (ZAMS) and terminal age main-sequence (TAMS) lines for low metallicity limit ($Z=0.008$); for Solar metallicity ($Z=0.014$) and for high metallicity limit ($Z=0.040$) as given by \citet[PARSEC models,][]{Bressan12} since these models cover the full ranges of stellar masses of the current sample and their internal physics were updated recently \citep{Chen15}. Why these upper and lower $Z$ limits were chosen is demonstrated in Fig. 1, where corresponding $Z$ values were marked just above the histogram distribution. 

\begin{figure}[ht]
\begin{center}
\includegraphics[scale=1, angle=0]{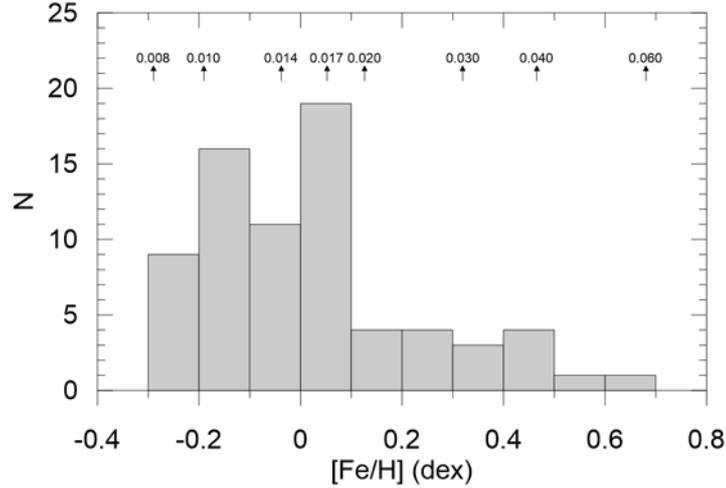}
\caption[] {Metallicity [Fe/H] distribution of detached eclipsing spectroscopic binaries based on 72 stars given in Table 2. The $Z$ values used by pre-computed PARSEC models \citep{Bressan12} are marked above for comparison.}
\end{center}
\end{figure}

The ZAMS line indicates the beginning of the main-sequence phase, while the TAMS indicates the end of the main-sequence phase, when hydrogen is exhausted ($X=0$) in the centre of the core. Main-sequence stellar lifetime is a strong function of $M$, in that massive stars have shorter lifetimes than less massive stars. Therefore, it is not possible to fix an age and say all the stars younger than this age are main-sequence. It is a known fact that the field main-sequence stars have different metallicities and different ages. Thus, all stars located between ZAMS and TAMS are considered on the main-sequence.
 
Because $M$ and $R$ are the most reliable and directly accessible observational parameters among the other parameters such as $L$, and $T_{eff}$ from the simultaneous solutions of radial velocity and light curves of detached eclipsing spectroscopic binaries, we have chosen to use the $\log M - \log R$ diagram to identify main-sequence stars and differentiate probable pre and post main-sequence stars using the PARSEC models and according to metallicity limits $Z=0.008$, and $Z=0.040$, as indicated by the histogram distribution in Fig. 1. Fig. 2 shows the candidate stars rejected and the selected stars retained for our study sample, based on the constraints on metallicity and age using the ZAMS and TAMS lines from PARSEC models \citep{Bressan12}. We have identified 46 stars above the TAMS lines which could be considered probable non-main-sequence stars either evolved off the main sequence, most likely for massive ($M>1M_{\odot}$), or evolving towards the main-sequence, most likely for less massive ($M<1M_{\odot}$) stars. We identified one star (primary of TY Cra) below the ZAMS lines. After discarding 47 stars, the number of stars in the study sample is reduced to 510 stars, which all are main-sequence stars in the Solar neighbourhood and the Galactic disc, with a good probability of having an approximately Solar metallicity of $\langle{\rm [Fe/H]}\rangle=0$ dex, with a standard deviation $\sigma_{{\rm [Fe/H]}}=0.218$ dex.

\begin{figure}
\begin{center}
\includegraphics[scale=1, angle=0]{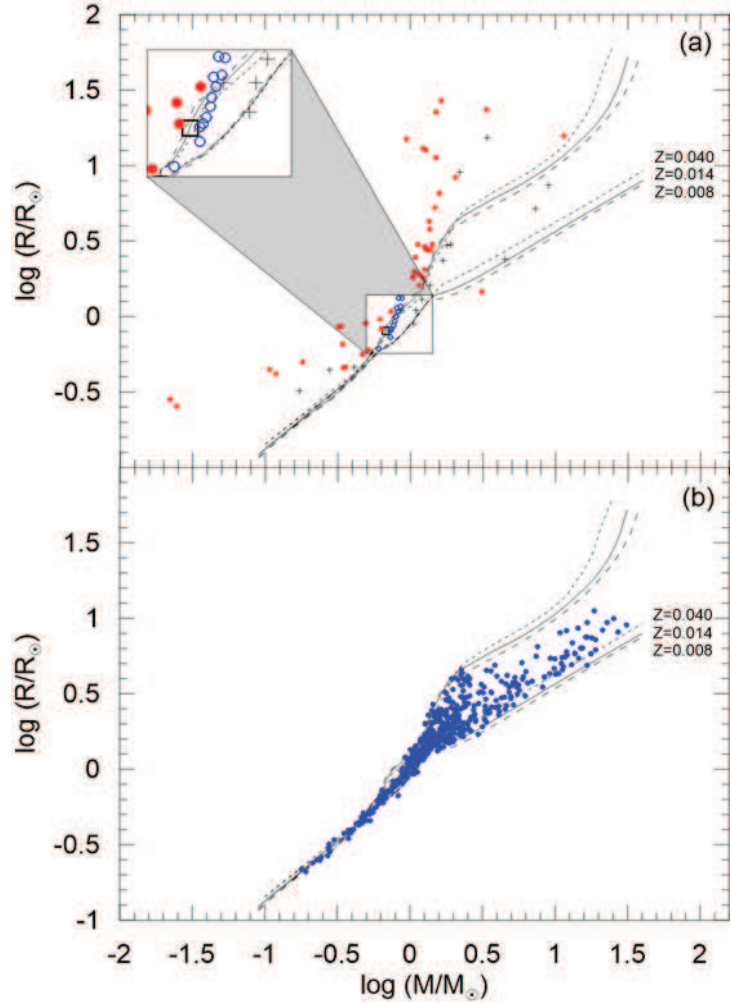}
\caption[] {(a) Identifying non-main-sequence stars (red dots above the TAMS lines), b) Study sample (blue dots) selected main-sequence stars within the limits of ZAMS and TAMS lines according to PARSEC models \citep{Bressan12}. Metallicities were indicated at the end of the ZAMS lines. Metallicities are not marked for the TAMS lines but same line style indicates same metallicity. Empty circles (a) show Globular cluster members. Plus sign marks the stars with relative $M$ and $R$ errors bigger than 15\%. Densest part of distribution is expanded for clarity. Only one star is found below ZAMS (a).}
\end{center}
\end{figure}

We found only one binary (HD 187669) of both components appear to be evolved of the main-sequence in the sub-sample with metallicities (Fig. 1). So, we may consider that 70 out of 509 ($\sim 14\%$) stars in the study sample have metallicity measurements. Thus, we may assume that the metallicity distribution of the study sample is similar to the distribution displayed in Fig. 1. 

\subsubsection{Special case, primary of T-Cyg1-12664}
The primary of T-Cyg1-12664 (KIC 10935310) is a peculiar star. It is oversized, spotted and hotter than the stars in its mass range \citep{Iglesias-Marzoa17}, while its secondary is a cool star near the mass boundary for fully convective stars obeying the mass luminosity relation of low mass stars. Therefore, we also removed the primary of T-Cyg-12664 from the study sample, thus the number of stars in the study sample reduced to 509.

\subsubsection{Accuracies in the study sample}
There were 268 stars in the list of \citet{Eker15} who last calibrated the classical MLR. The number of main-sequence stars in this study is increased to 509. That is, the number of stars are almost doubled, where 345 stars have accuracies up to 3\%, 88 stars have accuracies 3-6\%, and 76 stars have accuracies 6-15\% for both $M$ and $R$. Deselected stars in the preliminary sample (77 stars) and selected stars (the study sample) and their $M$ and $R$ accuracies (3\%, 3-6\% or 6-15\%) are indicated in the last column of the Table 1. Why the present data were not limited to 3\% will be discussed later. Perhaps the most important aspect of the present sample is the extension of the mass range towards the low-mass stars. The low mass limit in the present sample is $0.179M_{\odot}$, while previously it was $0.38M_{\odot}$ \citep{Eker15}. Thus, $0.179\leq M/M_{\odot}\leq 31$ is the largest and most numerous mass ranges established for main sequence stars with accurate absolute parameters and an estimated metallicity distribution ($0.008\leq Z\leq 0.040$).

\section{Calibrations}
\subsection{The Classical MLRs}

Mass-luminosity distribution of the present study sample of 509 stars is displayed in Fig. 3. Definitely, it is more crowded and better constrained than the distribution displayed by \citet{Choi16}, who plotted DEBcat stars by \citet{Southworth15} and stars from the list of \citet{Torres10} \citep[see Fig. 21 in][]{Choi16}. Considering that the preliminary sample of this study already included components of the same detached eclipsing binaries given by \citet{Southworth15} and \citet{Torres10}, the constraints applied in this study successfully gave a narrower distribution when compared to theirs. For example, the distribution given by \citet{Choi16} suggests that their list includes components of detached binaries from globular clusters and the Magellanic Clouds. We eliminated such stars and allowed only Solar neighbourhood stars. Further, our study's sample stars have been additionally confirmed to be main-sequence stars, based on applying the theoretical ZAMS and TAMS lines given by PARSEC models \citep{Bressan12}, and to have metallicity limits within $0.008\leq Z\leq 0.040$.

\begin{figure*}
\begin{center}
\includegraphics[scale=0.6, angle=0]{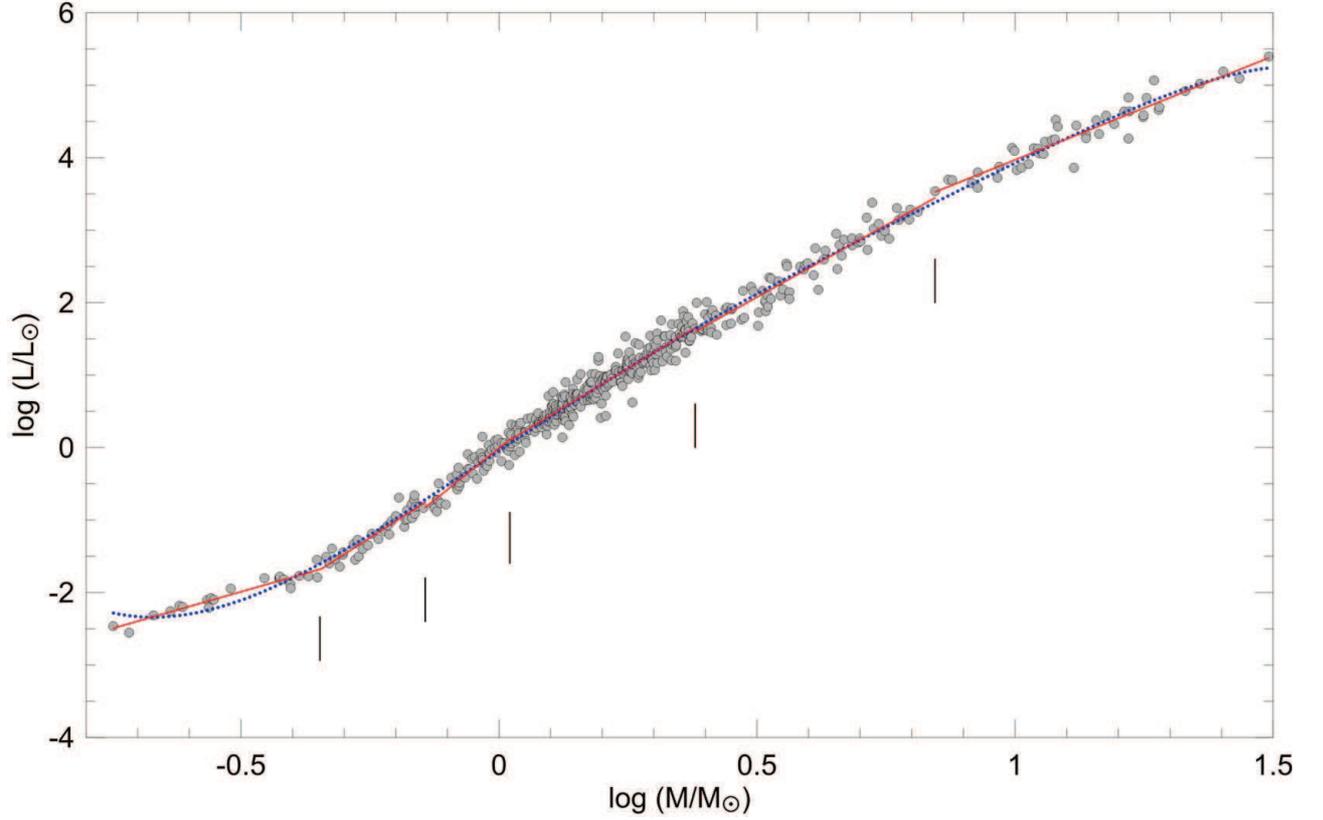}
\caption[] {The luminosity ($\log L/L_{\odot}$) distribution of the study sample (509 stars). Dotted (blue) line is a sixth degree polynomial, solid (red) lines are classical MLRs (linear fits) and the vertical lines are the break points separating mass domains where the linear lines were fit.}
\end{center}
\end{figure*}

In earlier studies, when the number of data was limited, a single MLR function could fit the existing data on a $\log M-\log L$ diagram. With the present study sample, however, in the full range of masses $0.179\leq M/M_{\odot}\leq 31$, as shown in Fig. 3, it is clearly evident that a physically meaningful simple function would not adequately fit to the luminosities for all 509 main-sequence stars. One way to describe the present data is to use multiple lines ($\log L=a\log M + b$, where $a$, $b$ are coefficients, to be determined) or a polynomial with $n+1$ number of coefficients, where $n$ is the degree of the polynomial, which would fit the present data set at best. In the first case, the number of linear lines, in the second case the degree of polynomial is arbitrary. 

We applied an F-test in order to see which method is feasible and physically more meaningful. Starting from $n=1$ up to $n=6$, we produced six polynomials fitting the full range of the data displayed in Fig. 3 by the least squares method. Polynomials were evaluated and compared to the six-piece linear functions (classical MLRs), where the piece of polynomial and best fitting line were computed at 500 random mass points between predetermined break points shown in Fig. 3. The break points separating the mass domains (ranges) could we chosen arbitrarily. Instead, we chose to maintain the previously determined three break points and four mass domains established by \citet{Eker15}, and added two more break points together with the extended mass range. Thus, the number of mass domains in our present study has increased to six. Our study's requirements for, and the physical significance of these break points will be discussed later. Table 3 displays the F-test results, where column one gives the domain name. The second column indicates the degree of the polynomials which were evaluated and compared to the linear function (linear MLR) for each domain. In Table 3, the F-value and probability ($p$) of each test is given in column three and four, while correlation coefficients ($R^2$) and standard deviations ($\sigma$) are given in columns six and seven. Note that, ($R^2$) values do not indicate correlations between the data and the fitting functions, but ($R^2$) values show how the linear MLR and polynomial MLRs correlates. Similarly, ($\sigma$) do not indicate standard deviation of data, but instead shows the standard deviation between the linear MLR and the polynomial MLRs, which were computed at 500 mass points randomly chosen within each domain. For a successful test, the F values must be smaller but larger than the $p$ values, where ($R^2$) must be close to one and ($\sigma$) must be minimal. 

\begin{table}
\setlength{\tabcolsep}{2pt}
\centering
\caption{Comparing full range polynomial fits to the linear MLR in each domain.}
\begin{tabular}{|cccccc|}
\hline
Domain & Degrees & $F$ & $p$ & $R^2$ & $\sigma$ \\
\hline
1 & 2 & 18.413 & 0.005 & 0.032 & 0.350 \\
  & 3 & 16.050 & 0.006 & 0.412 & 0.191 \\
  & 4 &  9.139 & 0.013 & 0.759 & 0.122 \\
  & 5 &  8.892 & 0.014 & 0.706 & 0.135 \\
  & 6 &  8.318 & 0.015 & 0.730 & 0.130 \\
\hline
2 & 2 & 25.012 & 0.003 & 0.425 & 0.222 \\
  & 3 & 22.112 & 0.004 & 0.584 & 0.189 \\
  & 4 & 12.225 & 0.009 & 0.785 & 0.136 \\
  & 5 &  5.499 & 0.028 & 0.838 & 0.118 \\
  & 6 &  5.379 & 0.029 & 0.839 & 0.117 \\
\hline
3 & 2 & 23.632 & 0.003 & 0.612 & 0.176 \\
  & 3 & 12.532 & 0.009 & 0.722 & 0.149 \\
  & 4 &  7.121 & 0.019 & 0.756 & 0.139 \\
  & 5 &  6.038 & 0.025 & 0.763 & 0.137 \\
  & 6 &  5.786 & 0.026 & 0.765 & 0.137 \\
\hline
4 & 2 & 14.595 & 0.007 & 0.896 & 0.144 \\
  & 3 & 19.867 & 0.004 & 0.894 & 0.146 \\
  & 4 &  9.960 & 0.012 & 0.898 & 0.143 \\
  & 5 &  3.264 & 0.058 & 0.900 & 0.141 \\
  & 6 &  3.418 & 0.054 & 0.900 & 0.141 \\
\hline
5 & 2 & 13.091 & 0.062 & 0.897 & 0.176 \\
  & 3 &  8.824 & 0.014 & 0.888 & 0.184 \\
  & 4 &  9.461 & 0.013 & 0.885 & 0.186 \\
  & 5 &  3.610 & 0.050 & 0.895 & 0.177 \\
  & 6 &  3.879 & 0.046 & 0.895 & 0.178 \\
\hline
6 & 2 & 12.598 & 0.009 & 0.845 & 0.182 \\
  & 3 &  6.508 & 0.022 & 0.873 & 0.164 \\
  & 4 &  6.383 & 0.023 & 0.871 & 0.166 \\
  & 5 &  4.262 & 0.040 & 0.880 & 0.160 \\
  & 6 &  5.236 & 0.030 & 0.877 & 0.161 \\
\hline
\end{tabular}%
\label{tab:addlabel}%
\end{table}%

Looking at each domain in Table 3, the F-values get smaller and become saturated as the degree of the polynomials increases. Since the F-values get smaller while being bigger than the $p$ values, a higher degree polynomial is better at representing the data than the polynomial with a lower degree. Although a higher degree polynomial is more eligible fitting data, there must be a point to stop increasing its degree because the standard deviation reaches its optimal value and starts not changing anymore by increasing the degree of the polynomial. According to Table 3, a polynomial of a sixth degree (implied by the F-test in domains 1, 2 and 3) or a polynomial of a fifth degree (implied by the F-test in domains 4, 5 and 6) is equally likely to be used instead of six-piece classical (linear) MLR. This result of Table 3 is also confirmed in Fig. 3 that the sixth degree polynomial and linear MLRs are not differentiable (appear same) especially in domains four and five, but the data appear to be represented better by linear MLRs for the other domains; the difference between the line and the polynomial is especially noticeable in the first and in the last domains. Although, for representing the data, a sixth degree (or fifth degree) polynomial is equal to the linear lines (classical MLRs), such a polynomial does not produce physically meaningful coefficients as the linear lines, where the first coefficient is the power of $M$ for the classical MLR ($L \propto M^{\alpha}$). Therefore for this study we preferred the six-piece classical linear MLR rather than single-piece fifth or sixth degree polynomial MLR to represent the observed mass-luminosity distribution displayed in Fig. 3.    

The mass-luminosity diagram of the study sample is shown in Fig. 4.  The best fitting linear lines, which we considered as classical MLRs, and $1\sigma$ limits in the mass ranges separated by the break points are shown in the six panels below. The coefficients and errors in the coefficients determined by the least squares method and according to standard error analysis techniques of the least squares are displayed in Table 4, where columns indicate the name of the domain, number of stars, mass range of the domain, classical MLRs (linear equation on $\log M-\log L$ plane, where the numbers in the parenthesis are the errors of the coefficients), correlation coefficients ($R^2$), standard deviations ($\sigma$) and the inclination of the line ($\alpha$), which is the power of $M$ if MLR is expressed as $L \propto M^{\alpha}$.

\begin{figure*}
\begin{center}
\includegraphics[scale=0.8, angle=0]{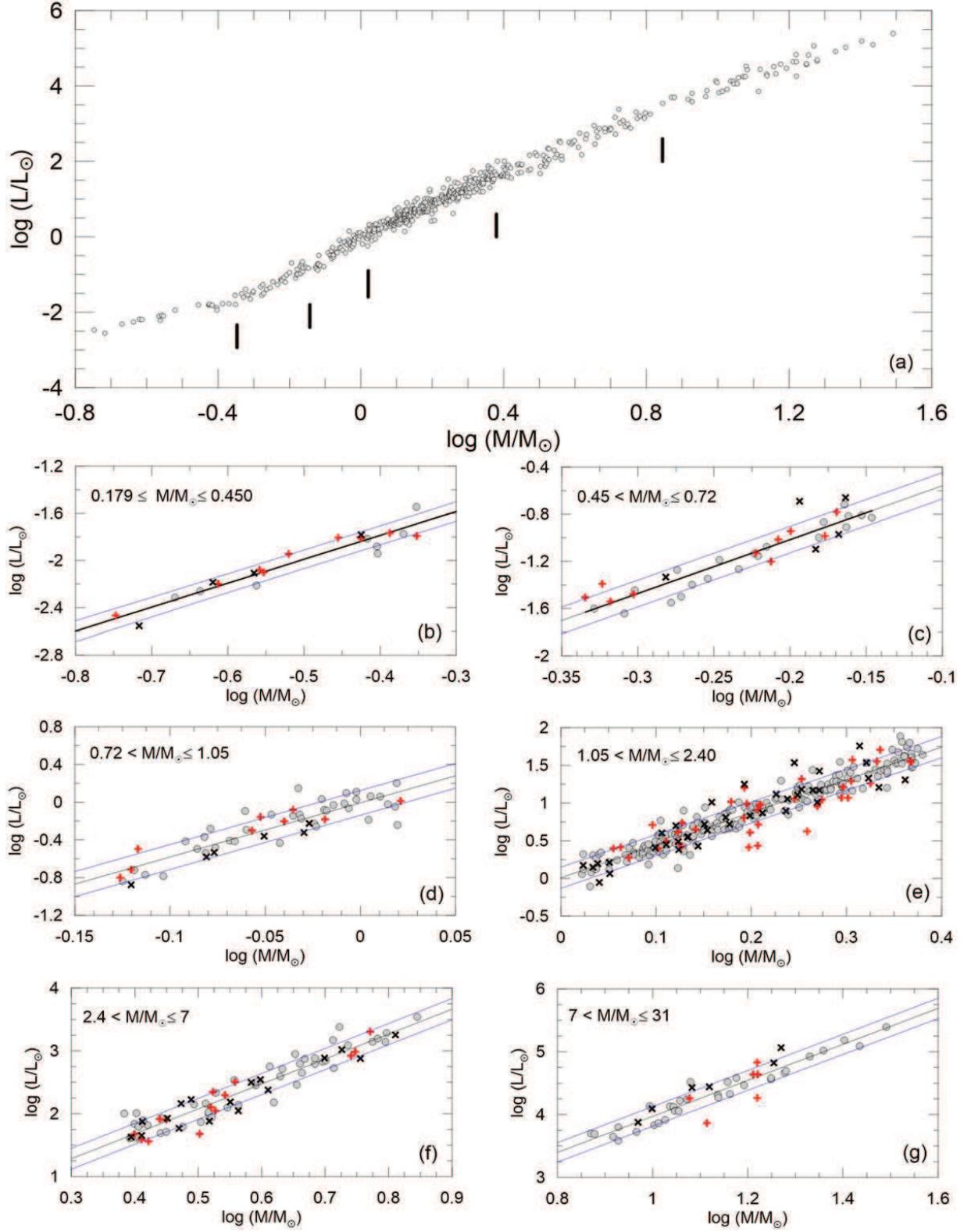}
\caption[] {The $M-L$ diagram of the study sample (a) where vertical dashes corresponding to the break points separating the ultra low-mass, very low-mass, low mass, intermediate mass, high mass and very high-mass domains. The lower six panels (b, c, d, e, f, g) show best fitting lines and $1\sigma$ limits in those domains. Data accuracy is indicated as (o) very accurate $<3\%$, (+) accurate (3-6\%), ($\times$) less accurate (6-15\%).}
\end{center}
\end{figure*}

\begin{table}
\setlength{\tabcolsep}{5pt}
\centering
\caption{Classical MLRs for main-sequence stars at various mass domains.}
\begin{tabular}{lclcccc}
\hline
Domain            & $N$  & Mass Range & Classical MLR & $R^2$ & $\sigma$     & $\alpha$ \\
\hline
Ultra low-mass     &  22  & $0.179<M/M_{\odot}\leq0.45$  & $\log L=2.028(135)\times\log M-0.976(070)$ & 0.919 & 0.076 & 2.028 \\
Very low-mass      &  35  & $0.45< M/M_{\odot}\leq0.72$  & $\log L=4.572(319)\times\log M-0.102(076)$ & 0.857 & 0.109 & 4.572 \\
Low mass           &  53  & $0.72< M/M_{\odot}\leq1.05$  & $\log L=5.743(413)\times\log M-0.007(026)$ & 0.787 & 0.129 & 5.743 \\
Intermediate mass & 275  & $1.05< M/M_{\odot}\leq2.40$  & $\log L=4.329(087)\times\log M+0.010(019)$ & 0.901 & 0.140 & 4.329 \\
High mass         &  80  & $2.4 < M/M_{\odot}\leq 7$    & $\log L=3.967(143)\times\log M+0.093(083)$ & 0.907 & 0.165 & 3.967 \\
Very highmass    &  44  & $7< M/M_{\odot}\leq31$       & $\log L=2.865(155)\times\log M+1.105(176)$ & 0.888 & 0.152 & 2.865 \\
\hline
\end{tabular}%
\label{tab:addlabel}%
\end{table}%

According to the statistics given in Table 4, the power of $M$ is smallest ($\alpha=2.028$) in the ultra low mass domain ($0.179\leq M/M_{\odot}\leq 0.45$), then it increases and reaches its maximum ($\alpha=5.743$) in the low mass domain ($0.72<M/M_{\odot}\leq 1.05$), then starts decreasing where the lowest value ($\alpha=2.868$) is in the very high mass domain ($7<M/M_{\odot}\leq 31$). The standard deviation, on the other hand, characterizing the width of the distribution on the $\log M-\log L$ diagram appears to be increasing from ($\sigma=0.076$) in the ultra low-mass to ($\sigma=0.165$) in the high mass domains. Although $\sigma=0.152$ in the very high mass domain is smaller than $\sigma=0.165$ in the high mass domains, it is bigger than $\sigma=0.140$ in the intermediate mass domain, which may be considered the central or main part of the MLR described in this study. The correlation coefficients all indicate good consistency. Indeed, the correlations obtained support our study's choice to use a six piece MLR, because it is more practical and physically more meaningful than a single piece polynomial.  	

\subsection{MRR and MTR} 
The distribution of observational radii and effective temperatures of the present study sample are displayed in Fig. 5. Except for the low mass end, general appearance is similar to the diagrams shown by \citet{Eker15}. Thus, the $\log M - \log R$ diagram is still discouraging to define a unique MRR, even though it is possible mathematically by various methods; e.g. the least squares method. The width of the distribution is not uniform throughout the full mass range; thus it is very dissimilar to the distribution of luminosities on the $\log M - \log L$ diagram (Fig. 3). On the other hand, the distribution on the $\log M - \log T_{eff}$ diagram roughly resembles the distribution on the $\log M - \log L$ diagram. However, because of the wavy appearance of the distribution for the range of masses $M<1.5 M_{\odot}$, it does not appear possible to find a single, smooth and adequately fitting function capable of representing both the wavy part and the rest of the temperatures. 

\begin{figure*}
\begin{center}
\includegraphics[scale=0.5, angle=0]{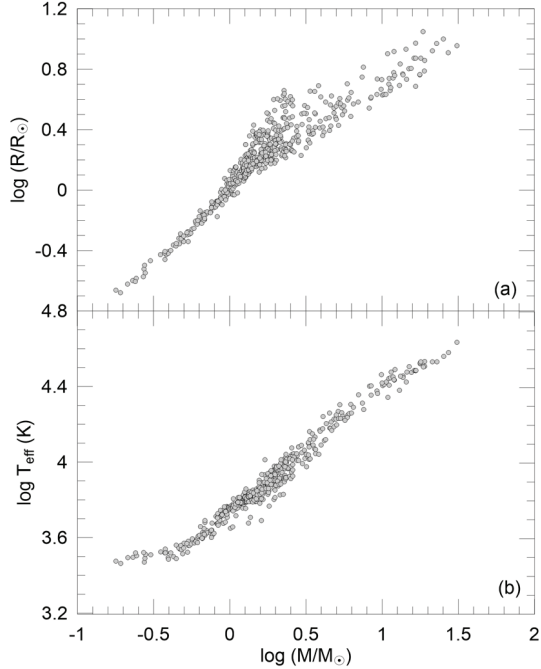}
\caption[] {The mass-radius (a) and the mass-effective temperature (b) diagrams of the study sample.}
\end{center}
\end{figure*}

Since neither diagram appears amenable to defining a single MRR or MTR, we have chosen to study the interdependence between the luminosity $L$ and other published parameters, including $M$, $R$ and $T_{eff}$. The calibrated MLR functions provide $L$ as a function of $M$ for the full range of masses in this study. Similarly, the Stefan-Boltzmann law requires that published $R$ and $T_{eff}$ cannot be independent of $L$ as well as $M$. Therefore, having the parameter $M$ on the horizontal axis, and the same stars on all diagrams ($\log M - \log L$, $\log M - \log R$, and $\log M - \log T_{eff}$), the interdependence of the three diagrams is valid at each value of $M$ as well as over the full range of masses.

The interdependence, however, allows one to choose a particular part of $\log M - \log R$ diagram that is able to be fit by a smooth MRR function, and to then compute the $R(M)$ function for the rest by using the MLR and MTR functions. The same is true with the $\log M - \log T_{eff}$ diagram. The smooth part of it could be chosen to calibrate an empirical MTR and then, for the rest the $T_{eff}(M)$ could be computed from the MLR and MTR functions. This approach not only solves the problem of defining MRR and MTR for the full range of masses, but also gives us an opportunity to confirm or reject pre-determined MLR and MRR, from which $T_{eff}(M)$ were computed. Vice versa to confirm or reject the pre-determined MLR and MTR, from which $R(M)$ were computed. As a result, the data on the broad band-like part of the $\log M - \log R$ diagram, and the data on the wavy part of the $\log M - \log T_{eff}$ diagram will be useful for confirming the initially determined and interrelated MLR, MRR and MTR functions.
   
Examining $\log M - \log R$ diagram (Fig. 5a), one can see the distribution of stellar radii for the stars with $M\leq 1.5 M_{\odot}$ ($\log M/M_{\odot}\leq 0.176$) is narrow and smooth. Thus, this is the part of that diagram most eligible to define an empirical MRR. On the other hand, stars with ($M>1.5 M_{\odot}$) on the $\log M - \log T_{eff}$ (Fig. 5b) are smoother than the rest of data, thus it is most eligible to define an empirical MTR. 

\subsubsection{Calibrating empirical MRR for the range $0.179\leq M/M_{\odot}\leq1.5$}

An empirical MRR was fixed using the published radii of stars having $M\leq 1.5 M_{\odot}$ in the present sample by the least squares method. $M$ and $R$ values of main-sequence stars with masses $M\leq1.5M_{\odot}$ are both sufficiently small that fitting process and trials do not require a logarithmic scale. After many trials with different functions, we found that the published radii in the range $0.179\leq M/M_{\odot}\leq 1.5$ are best represented by a quadratic function on a linear scale. That is, a quadratic function was found to produce the best fit when compared to linear and cubic equations. The coefficients and errors in the coefficients (given in parenthesis) determined by the least squares method and according to standard error analysis techniques of the least squares are displayed in Table 5, where the columns indicate the mass range, number of stars in the mass range, MRR function, correlation coefficient ($R^2$) and standard deviation ($\sigma$). A MRR function has not been determined for the mass range $1.5 < M/M_{\odot}\leq 31$. Instead, since $R(M)=R(L, T_{eff})$ the values and standard deviation shown were computed from the MLR and MTR.

\begin{table}
\setlength{\tabcolsep}{5pt}
\centering
\caption{Calibrations of empirical MRR at $M/M_{\odot}\leq 1.5$ and empirical MTR at $M/M_{\odot}>1.5$.}
\begin{tabular}{lcccc}
\hline
Mass range & Number & Emprical MRR & $R^2$ & $\sigma$ \\
\hline
$0.179\leq M/M_{\odot}\leq 1.5$ & 233 & $R=0.438(098)\times M^2+0.479(180)\times M+0.075(479)$ & 0.867 & 0.176 \\
$1.5< M/M_{\odot}\leq31$    & 276 & To be computed from MLR and MTR using $L=4\pi R^2\sigma T^4$ & $--$ & 0.787 \\
\hline
Mass range & Number & Emprical MTR & $R^2$ & $\sigma$ \\
\hline
$0.179\leq M/M_{\odot}\leq 1.5$ & 233 & To be computed from MLR and MRR using $L=4\pi R^2\sigma T^4$ & $--$ & 0.025 \\
$1.5< M/M_{\odot}\leq31$    & 276 & $\log T_{eff}= -0.170(026)\times (\log M)^2+0.888(037)\times \log M+3.671(010)$ & 0.961 & 0.042\\
\hline
    \end{tabular}%
  \label{tab:addlabel}%
\end{table}%

Figure 6a displays the best fitting MRR function (solid line) and the data in the mass range $0.179\leq M/M_{\odot}\leq 1.5$. The two parallel lines, one above and one below, indicate the one sigma limit. One may notice that radii of main-sequence stars with masses smaller than Solar mass are well contained with the one sigma limit. It is evident that the radii of main-sequence stars with masses smaller than $1M_{\odot}$ are well contained with the one sigma limit. Stellar evolution becomes effective to disperse observed radii towards the high mass end, so that the distribution of stars overflow the one sigma limit if $M>\sim 1.15 M_{\odot}$.    

\subsubsection{Calibrating empirical MTR for the range $1.5< M/M_{\odot}\leq31$}

An empirical MTR was determined from the published effective temperatures of stars with $M>1.5 M_{\odot}$ by the least squares method. Linear, quadratic and cubic functions were tried on both logarithmic and linear scales. We find that the published effective temperatures in the mass range $1.5<M/M_{\odot}\leq31$, are best represented by a single quadratic function on the $\log M - \log T_{eff}$ plane. The coefficients and errors in the coefficients (given in parenthesis) determined by the least squares method and according to standard error analysis techniques of the least squares are given in Table 5, in the same format except that the calibrated function [$T_{eff} (M)$] is given in the logarithmic scale. An open form of $T_{eff} (M)$ has not been assigned for the low mass stars in the mass range of $0.179\leq M/M_{\odot}\leq1.5$. Instead, since $T_{eff}(M)=T_{eff}(L, R)$, the values and standard deviation shown were computed from the MLR and MRR.   

The calibrated MTR function $[T_{eff}(M)]$ is shown together with the published temperatures in Fig. 6b on a logarithmic scale. The horizontal axis (masses) in Fig. 6a (upper panel), which are shown on a linear scale, continues from the point where it is cut, but in a logarithmic scale in Fig. 6b (lower panel). Therefore, the same numbers in the horizontal axis do not imply the same masses.

\begin{figure*}
\begin{center}
\includegraphics[scale=1, angle=0]{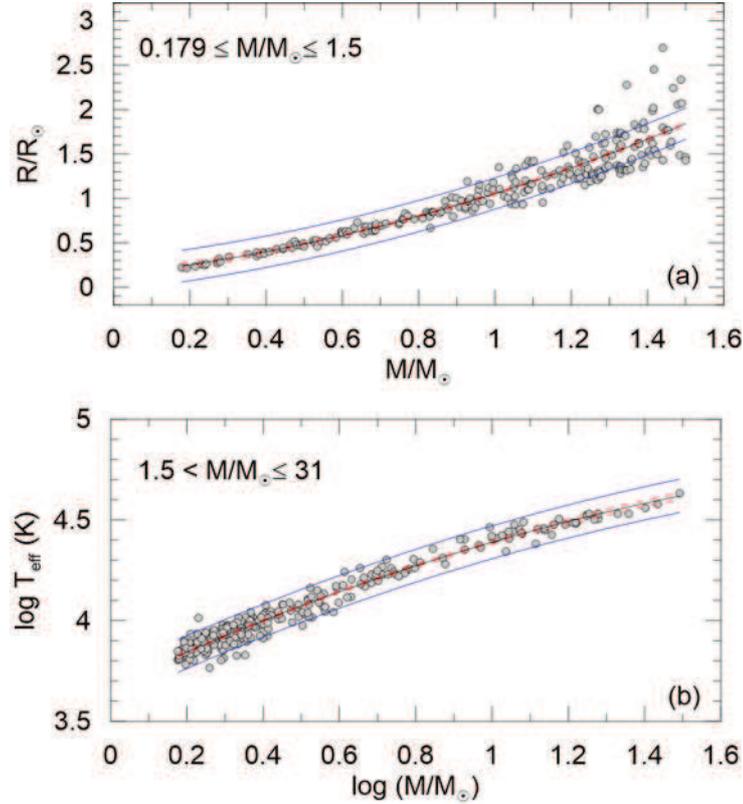}
\caption[] {(a) The best fitting MRR curve and $1\sigma$ limit (blue solid) for $M\leq 1.5 M_{\odot}$ and (b) The best fitting MTR curve and $1\sigma$ limit (blue solid) for $M>1.5M_{\odot}$. The dashed lines (red) show the confidence levels of the functions.}
\end{center}
\end{figure*}

\subsection{Interrelated MLR, MRR and MTR for the full range $0.179\leq M/M_{\odot}\leq31$}

The smoothness of the $R(M)$ function (solid line) in the range $0.179\leq M/M_{\odot}\leq1.5$ and the smoothness of the $T_{eff}(M)$ function (solid line) in the range $1.5< M/M_{\odot}\leq31$ ($0.176< \log (M/M_{\odot})\leq1.491$) is clearly evident in Fig. 6. Whereas a considerable number of the most massive stars overflow the one sigma limit in the upper diagram representing the radii and MRR function, almost all the data in the lower diagram in comparison, showing published temperatures and the MTR function, are well contained within the one sigma limit (Fig. 6).

For our study, we felt it important to obtain two partial smooth functions (MRR, MTR), one for radii and one for temperatures, covering the full range of main-sequence masses. This is necessary in order to determine interrelated MLR, MRR and MTR functions over the full range of masses of the present study sample of main-sequence stars.

\section{Discussions}
\subsection{Break points on MLR}
Calibrated MLR, MRR and MTR functions and observational data are displayed in Fig. 7. Vertical lines mark the positions of the break points. Between the break points, a MLR is classical; $L \propto M^{\alpha}$, where the power of $M(\alpha$) is a constant and unique to the domain. Unlike, MLR which were calibrated for the full range ($0.179\leq M/M_{\odot}\leq31$), MRR were calibrated only for the low mass stars ($M\leq1.5 M_{\odot}$) while MTR were determined for the high mass ($M>1.5 M_{\odot}$) region, so that the full range of masses were covered. Because MLR, MRR and MTR are interrelated, one is able to compute $R(M)$ and $T_{eff}(M)$ curves for the full range, where $T_{eff}(M)$ curve for the low mass stars ($M\leq1.5 M_{\odot}$) is computed from MLR and MRR, while $R(M)$ curve for the massive stars are computed from MLR and MTR according to the Stefan-Boltzmann law. 

\begin{figure*}
\begin{center}
\includegraphics[scale=1, angle=0]{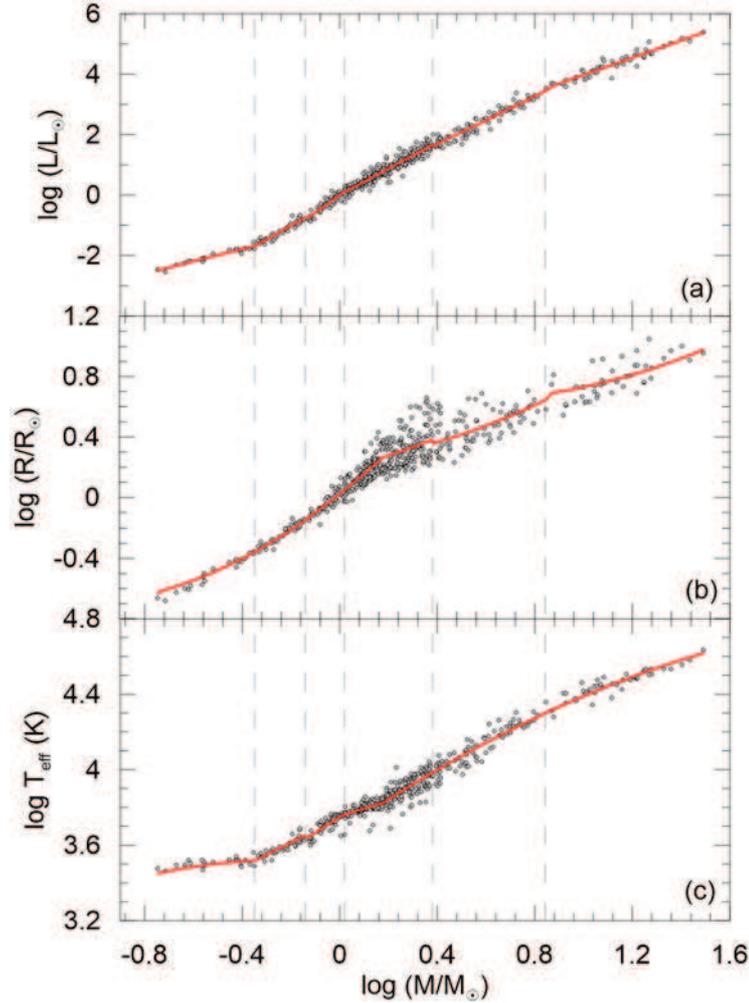}
\caption[] {The filled circles (data) are main-sequence $M-L$ (a), $M-R$ (b) and $M-T_{eff}$ (c) distributions. The solid lines are interrelated MLR (a), MRR (b) and MTR (c) functions. MTR at $M/M_{\odot}\leq 1.5$ and MRR at $M/M_{\odot}>1.5$ are computed from the other two according to the  Stefan-Boltzmann law. Vertical lines mark positions of break points on MLR and corresponding positions on MRR and MTR curves.}
\end{center}
\end{figure*}

Since the F-test in Section 3.1 indicated that there is no simple function able to replace a MLR function formed from six-linear lines following one after another over the full range of the mass-luminosity distribution, the break points are inevitable. The number and positions of the break points could have been arbitrarily, such that a MLR function is better represented by numerous lines as long as there is a statistically significant number of data between the successive break points. Despite the fact that existing data made us to think that the number and the positions of the break points are not really arbitrary. 

Break points in the six-piece MLR function indicate that the energy generation rate of main-sequence stars is not always a smoothly increasing function of $M$. Thus, there are certain mass-ranges where the power of $M$ is constant; $\log (L/L_{\odot})$ increases monotonically by the increase of $\log (M/M_{\odot})$. Then, there are break points on the mass axis where energy production changes abruptly. Based on the present study sample, there appears to us to be five non-arbitrary break points. 

It is interesting to note however, that among the three higher mass break points of $\sim 1.05$, $\sim 2.4$ and $\sim 7M_{\odot}$ found by \citet{Eker15}, at least one can be logically related to the p-p chain since ``The p-p chain is the main energy source for stars less massive than the Sun, whereas the CNO cycle becomes dominant for the stars more massive than the Sun. Thus, we surmise that the break point at $1.05M_{\odot}$ is just an indication of this change. There could be similar reasoning related to the efficiency of stellar energy production mechanisms at the other break points...'' already commented in \citet{Eker15}, who encouraged nuclear astrophysicists to further investigate the physical facts behind these break points. 

With the present study's sample of stars enlarged to include more low mass stars, we appear to have identified two more break points, those at $\sim 0.45M_{\odot}$ and $\sim 0.72M_{\odot}$. Both break points are clearly evident on both the mass-luminosity and the mass-temperature diagrams in Fig. 7a and Fig. 7c. Actually, the first break points in Fig. 7c are reflections of the break points in Fig. 7a because MRR function in this region is smooth and MTR curve was computed from MLR (top panel) and MRR (middle panel). The break points at $\sim 2.4M_{\odot}$ and $\sim 7M_{\odot}$, which are barely noticeable on the top panel, appear as jumps on the $R(M)$ curve in the middle panel of Fig. 7. These jumps too must be the reflections of the break points separating intermediate, high and very high mass domains on MLR. This is because, the $R(M)$ curve for the massive stars ($M>1.5M_{\odot}$) is computed from the curves of MLR (top panel) and MTR (bottom panel), where MTR is known to be continuous and smooth. Closer inspection of the data and the solid lines in the region $1.5\leq M/M_{\odot}\leq 31$ indicate that the data indeed follow the trend and breakpoints, despite a large scatter in radii. Note the sudden drop at $M\sim 2.4M_{\odot}$ ($\log M = 0.380$), and jump up at $M\sim 7M_{\odot}$ ($\log M = 0.845$).
 
Non-arbitrariness of the break points is also implied by theoretical mass-luminosity diagram drawn by theoretical TAMS and ZAMS lines. The break points drawn on Fig. 8, which shows the theoretical mass-luminosity diagram of PARSEC models \citep{Bressan12}, appear not at arbitrary locations but at uniquely described locations. The first break point ($M\sim 0.45M_{\odot}$) is on the location that the rate of increase of $L$ changes from a slower to a faster increase at $\log (M/M_{\odot})\sim -0.347$. Such a change is not noticeable elsewhere on the diagram. The point at $M\sim7 M_{\odot}$ is also unique. For the masses $M<7M_{\odot}$, luminosity of metal-rich stars are less than the luminosity of metal-poor stars. This fact changes the direction at $M\sim7M_{\odot}$, metal-rich stars becomes more luminous than the metal-poor stars for the masses $M>7M_{\odot}$. The break point at $M\sim 2.4M_{\odot}$ is also unique because for the stars $M>2.4M_{\odot}$, a TAMS of any metallicity within $0.008\leq Z\leq 0.040$ is higher than (more luminous) the ZAMS of the lowest metallicity. But for the stars $M<2.4M_{\odot}$, the TAMS of $Z=0.040$ stars appear to be very close to or perhaps below the ZAMS of $Z=0.008$ stars. The situation changes back again for the masses $M<1.05M_{\odot}$ which is the third break point. Notice that the width of the main-sequence begins to increase for the stars $M<1.05M_{\odot}$. The width begins to decrease again for masses $M<0.72M_{\odot}$, which is the second break point. We conclude therefore, that the break points at $\sim 0.45$, $\sim0.72$, $\sim 1.05$, $\sim 2.4$ and $\sim 7M_{\odot}$ coincide with special locations along the mass axis with respect to stellar energy generation. Further, the break points we have identified along the mass axis, which we were not aware of prior to this study, were discovered serendipitously from their appearance in the MLR, MRR, and MTR functions obtained in this study, as based on both the theoretical ZAMS and TAMS lines of PARSEC models \citep{Bressan12} and the Stefan-Boltzmann law.

\begin{figure*}
\begin{center}
\includegraphics[scale=0.8, angle=0]{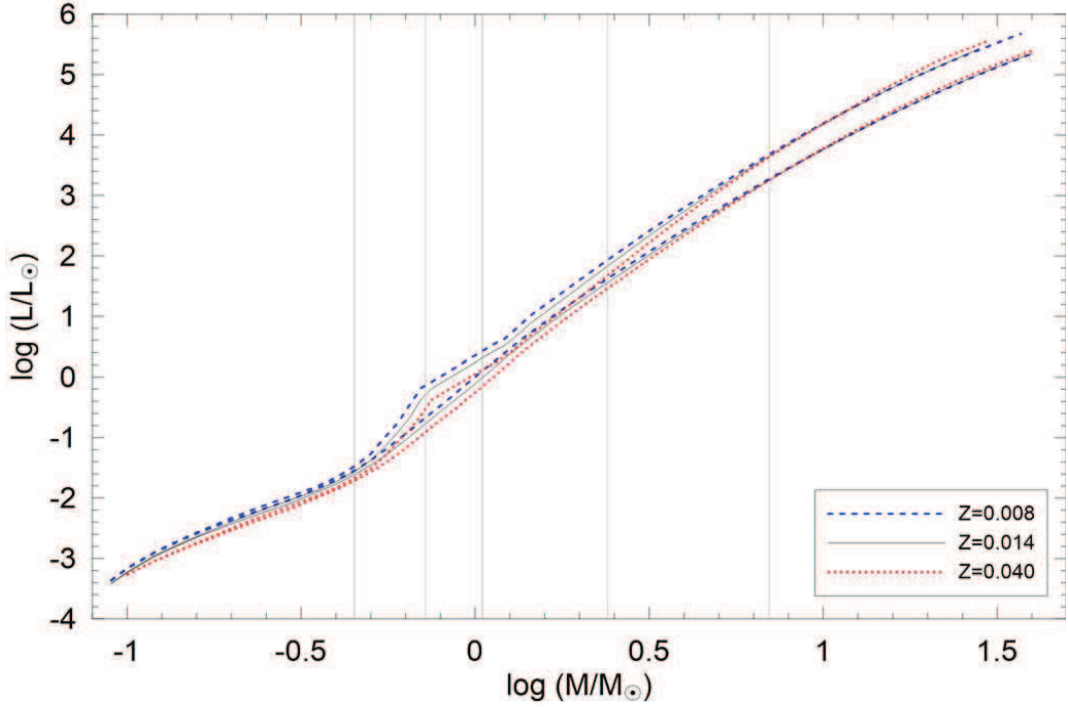}
\caption[] {Theoretical ZAMS and TAMS lines on $\log M - \log L$ diagram according to PARSEC models. Solar metallicity (black solid) and limiting metallicities (blue dashed $Z=0.008$, red dotted $Z=0.040$) are indicated. Vertical lines are the break points at $\sim 0.45$, $\sim 0.72$, $\sim 1.05$, $\sim 2.4$ and $\sim 7M_{\odot}$.}
\end{center}
\end{figure*}

Main-sequence stars are in a state of thermal equilibrium, such that their energy is radiated away from their surface at the same rate at which it is produced by nuclear reactions in their interior. Therefore, the change of luminosity as a function of $M$, as shown on the $\log M - \log L$ diagram, indicates a physical change in the mean energy generation rate as a function of $M$. Since energy generation rate is a direct result of the efficiency of p-p chain and/or CNO cycle, the exact underlying explanations of the break points is probably related to both the efficiency and the type of the nuclear reactions involved. The stars with $M<<0.45M_{\odot}$ are fully convective; thus, moving on mass axis while getting close to the first break point, a radiative region in the centre of the core develops and grows. So after the limit $M>0.45M_{\odot}$, the core gradually gets free of convection which brings fresh but cool material. Thus p-p chain reactions become more efficient. There could be two explanations for the break point at $0.72 M_{\odot}$. The first; p-p chain reactions must be improving not only by increase of pressure, density and temperature but also engaging various types of p-p chain, so each additional chain may contribute as an extra energy source in addition to already existing one. The second; although it would be inefficient like in the Sun, the CNO cycle reactions may start contributing for the stars $M>0.72M_{\odot}$. We already mentioned the possibility that CNO cycle reactions may dominate for the stars $M>1.05M_{\odot}$ and similar reasons related to the efficiency of stellar energy production mechanisms at the other break points. We encourage nuclear astrophysicists again to investigate the physical facts and reasoning behind all the break points.

\subsection{Locus values on MLR, MRR and MTR and comparing them to other determinations}

Interrelated MLR, MRR and MTR  are displayed in Fig. 9 together with the locus of main-sequence stars in the Solar vicinity and Galactic disc. The locus points are obtained by binning on the mass axis. Binning is fairly arbitrary, except that we used narrower bins for low mass stars and wider ones for higher mass stars. In addition, bins were optimized so as not to lose information and to be able to maintain the statistical significance of each bin. 

\begin{figure*}
\begin{center}
\includegraphics[scale=1, angle=0]{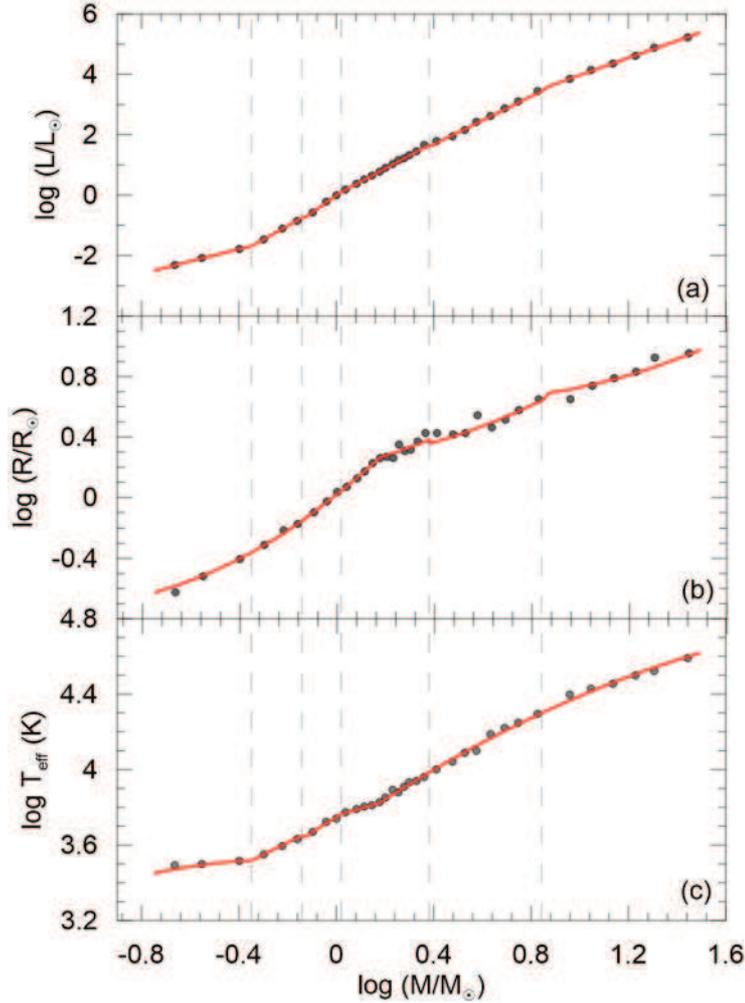}
\caption[] {Solid lines are interrelated MLR (a), MRR (b), and MTR (c). The filled circles are the locus points indicating the mean values from the current study sample. Vertical dashed lines indicate the break points.}
\end{center}
\end{figure*}

The mass ranges of the bins are listed in Table 6, where the number of stars in each bin is given in the second column. The next three columns display the mean values for the masses, radii and effective temperatures of the stars, which are grouped according to the mass ranges given in the first column (bins). A mean value is calculated as a simple arithmetic average, e. g. sum of masses divided by the number. Mean spectral types were also determined and listed in column six in Table 6. Published spectral types of the components listed in Table 1 were used in estimating the mean spectral type of the stars contained in each bin. Mean bolometric absolute magnitudes were calculated and listed in column seven according to Pogson's relation as:

\begin{equation}
M_{bol}-(M_{bol})_{\odot}=-2.5\log L/L_{\odot}
\end{equation}
where $(M_{bol})_{\odot}=4.74$ \citep{Cox00} is used. Mass to luminosity ratio is critical for extragalactic astronomers when modeling galaxies and/or searching for dark matter, while mean energy production for stars with different masses is of interest to nuclear astrophysicists, so both are included in the table.

\begin{table}
\setlength{\tabcolsep}{4pt}
\centering
\caption{Mean absolute parameters (locus points) of the present main-sequence stars sample.}
\begin{tabular}{cccccccccc}
\hline
Mass Range & $N$ & $\langle M \rangle$ & $\langle R \rangle$ & $\langle T_{eff} \rangle$ & $\log g$ & Spt   & $M_{bol}$ & $M/L$ & $L/M$\\
   &  & ($M_{\odot}$)    & ($R_{\odot}$)   & (K) &   (cgs) &     & (mag)  &  ($\odot$)  & (erg s$^{-1}$ gr$^{-1}$) \\
\hline
$0.16 < M \leq 0.25$ &  6 &  0.217 & 0.238 &  3105 & 5.020 & M4.5 & 10.55 & 45.7713 & 0.044 \\
$0.25 < M \leq 0.35$ &  5 &  0.281 & 0.302 &  3167 & 4.926 & M3.5 &  9.95 & 34.0746 & 0.057 \\
$0.35 < M \leq 0.45$ & 11 &  0.398 & 0.395 &  3282 & 4.844 & M3   &  9.21 & 24.4342 & 0.08 \\
$0.45 < M \leq 0.55$ & 12 &  0.503 & 0.489 &  3547 & 4.761 & M1.5 &  8.41 & 14.8043 & 0.13 \\
$0.55 < M \leq 0.65$ & 11 &  0.602 & 0.612 &  3924 & 4.644 & K9   &  7.49 &  7.5421 & 0.26 \\
$0.65 < M \leq 0.75$ & 14 &  0.690 & 0.675 &  4289 & 4.618 & K4.5 &  6.89 &  4.9844 & 0.39 \\
$0.75 < M \leq 0.85$ & 14 &  0.802 & 0.798 &  4663 & 4.539 & K3.5 &  6.16 &  2.9687 & 0.65 \\
$0.85 < M \leq 0.95$ & 19 &  0.907 & 0.935 &  5279 & 4.454 & K0   &  5.28 &  1.4882 & 1.30 \\
$0.95 < M \leq 1.05$ & 18 &  1.000 & 1.092 &  5498 & 4.362 & G5   &  4.76 &  1.0222 & 1.89 \\
$1.05 < M \leq 1.15$ & 21 &  1.093 & 1.176 &  5922 & 4.336 & G3   &  4.28 &  0.7156 & 2.70 \\
$1.15 < M \leq 1.25$ & 22 &  1.209 & 1.337 &  6181 & 4.269 & F8   &  3.82 &  0.5162 & 3.74 \\
$1.25 < M \leq 1.35$ & 43 &  1.300 & 1.482 &  6396 & 4.211 & F6   &  3.44 &  0.3941 & 4.91 \\
$1.35 < M \leq 1.45$ & 27 &  1.398 & 1.688 &  6495 & 4.129 & F4.5 &  3.09 &  0.3070 & 6.30 \\
$1.45 < M \leq 1.55$ & 23 &  1.502 & 1.823 &  6737 & 4.093 & F3   &  2.77 &  0.2443 & 7.91 \\
$1.55 < M \leq 1.65$ & 25 &  1.596 & 1.865 &  7110 & 4.100 & F1   &  2.49 &  0.2001 & 9.66 \\
$1.65 < M \leq 1.75$ & 12 &  1.702 & 1.818 &  7794 & 4.150 & A8   &  2.14 &  0.1554 & 12 \\
$1.75 < M \leq 1.85$ & 22 &  1.793 & 2.230 &  7560 & 3.995 & A6   &  1.83 &  0.1230 & 16 \\
$1.85 < M \leq 1.95$ & 18 &  1.894 & 2.035 &  8153 & 4.099 & A5   &  1.70 &  0.1153 & 17 \\
$1.95 < M \leq 2.05$ & 19 &  1.994 & 2.059 &  8606 & 4.111 & A4   &  1.44 &  0.0955 & 20 \\
$2.05 < M \leq 2.2$  & 17 &  2.139 & 2.353 &  8722 & 4.025 & A3   &  1.09 &  0.0744 & 26 \\
$2.2  < M \leq 2.4$  & 26 &  2.299 & 2.683 &  9154 & 3.943 & A2   &  0.60 &  0.0507 & 38 \\
$2.4  < M \leq 2.8$  & 20 &  2.573 & 2.653 & 10030 & 4.001 & A0   &  0.22 &  0.0402 & 48 \\
$2.8  < M \leq 3.2$  & 10 &  2.993 & 2.610 & 10999 & 4.081 & B9   & -0.14 &  0.0335 & 58 \\
$3.2  < M \leq 3.6$  & 12 &  3.362 & 2.657 & 12239 & 4.116 & B8   & -0.64 &  0.0236 & 82 \\
$3.6  < M \leq 4.0$  &  8 &  3.769 & 3.497 & 12588 & 3.927 & B7.5 & -1.36 &  0.0137 & 141 \\
$4.0  < M \leq 4.6$  &  8 &  4.310 & 2.911 & 15372 & 4.145 & B5.5 & -1.83 &  0.0101 & 190 \\
$4.6  < M \leq 5.2$  &  9 &  4.916 & 3.287 & 16576 & 4.096 & B4.5 & -2.42 &  0.0067 & 288 \\
$5.2  < M \leq 6.0$  &  8 &  5.587 & 3.797 & 17677 & 4.027 & B3.5 & -3.01 &  0.0044 & 437 \\
$6.0  < M \leq 8.0$  &  7 &  6.716 & 4.460 & 19729 & 3.967 & B2.5 & -3.84 &  0.0025 &  779 \\
$8.0  < M \leq 10 $  &  7 &  9.083 & 4.488 & 25057 & 4.092 & B2   & -4.89 &  0.0013 & 1516 \\
$10   < M \leq 12 $  & 11 & 11.143 & 5.497 & 26685 & 4.005 & B1   & -5.61 &  0.0008 & 2386 \\
$12   < M \leq 15 $  &  8 & 13.702 & 6.186 & 28583 & 3.992 & B0.5 & -6.16 &  0.0006 & 3235 \\
$15   < M \leq 18 $  &  8 & 16.886 & 6.825 & 31579 & 3.998 & O9.5 & -6.81 &  0.0004 & 4761 \\
$18   < M \leq 24 $  &  5 & 20.163 & 8.454 & 33220 & 3.889 & O8   & -7.49 &  0.0003 & 7492 \\
$24   < M \leq 32 $  &  3 & 27.835 & 9.037 & 39067 & 3.971 & O6   & -8.34 &  0.0002 & 11860\\
\hline
\end{tabular}%
\end{table}%

The classical MLRs for each domain as separated by break points (vertical lines), follow a smooth and continuous $R(M)$ curve for the low-mass stars ($M\leq1.5M_{\odot}$), and a smooth and continuous $T_{eff}(M)$ curve for the high mass stars ($M>1.5M_{\odot}$). The MLRs are given even clear when the locus points are added, as in Fig. 9. How good the wavy part of $T_{eff}(M)$ curve ($M\leq1.5$) and the broad part of the radius distribution ($M>1.5$) by $R(M)$ curve were supported by the data (locus) is clearly noticeable. The scatter of radii with the massive stars ($M>1.5$) in the middle panel indicates that the statistics is rather insufficient due to the results of the stellar evolution.

Calibrated MLR of this study is compared to other determinations in Fig. 10. Apparently, a very narrow region $1<M/M_{\odot}<1.5$ ($0<\log(M/M_{\odot})<0.18$) is the part that all other determinations agree. The disagreements are usually at both the lower and higher ends of the mass range. The two regions shown in squares are zoomed into for a better comparison. The difference between the present MLR and the previous MLR by \citet{Eker15} is clearly visible at the low mass end $M/M_{\odot}<0.45$. The MLR function of \citet{Malkov07} is a quadratic function of $\log M$ to provide $\log L$ within the limits $-2<\log (M/M_{\odot})<1.5$ (from 0.63 to 31.6$M_{\odot}$). \citet{Demircan91} gives two-linear equations valid for the mass range from 0.1 to 18.1$M_{\odot}$, the first one is a single line and the other is a two-piece line broken at $M=0.7M_{\odot}$. \citet{Demircan91} models overestimates the luminosities of low-mass stars $M<1M_{\odot}$ down to $\sim0.2$. The luminosities of massive stars ($M>1M_{\odot}$), on the other hand, appear to be slightly underestimated up to $\sim15M_{\odot}$. \citet{Malkov07} models give luminosities similar to the present study and \citet{Eker15} models up to $\sim 6M_{\odot}$, then appears deviate by giving slightly lower luminosities for the masses $M>6M_{\odot}$.

\begin{figure*}
\begin{center}
\includegraphics[scale=0.6, angle=0]{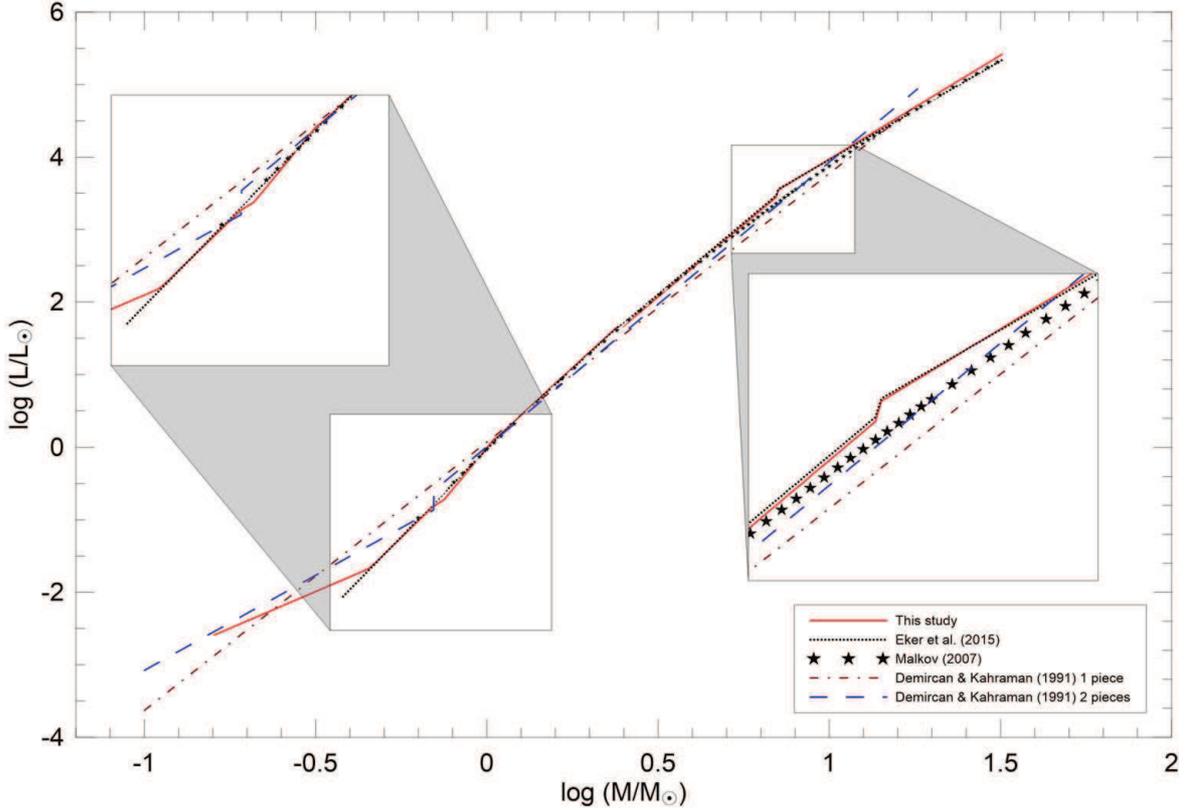}
\caption[] {Comparing calibrated MLR of this study to previous determinations, with two regions of interest enlarged for clarity.}
\end{center}
\end{figure*}

The calibrated MRR of this study is compared to other determinations in Fig. 11. The most recent MRR function is given by \citet{Malkov07}, which is a cubic function of $\log M$ to provide $\log R$ within the limits $-2<\log(M/M_{\odot})<1.5$. \citet{Demircan91} also gave two-linear equations valid for from 0.1 to 18.1$M_{\odot}$; the first one is a single line and the other is a two-piece broken line with a break point at $M=1.6M_{\odot}$. The most significant difference between the present study and the other three is that the others overestimate radii for stars of $M<1M_{\odot}$ and underestimate radii for stars of $M>1M_{\odot}$ up to roughly $\sim10M_{\odot}$. 

\begin{figure*}
\begin{center}
\includegraphics[scale=0.6, angle=0]{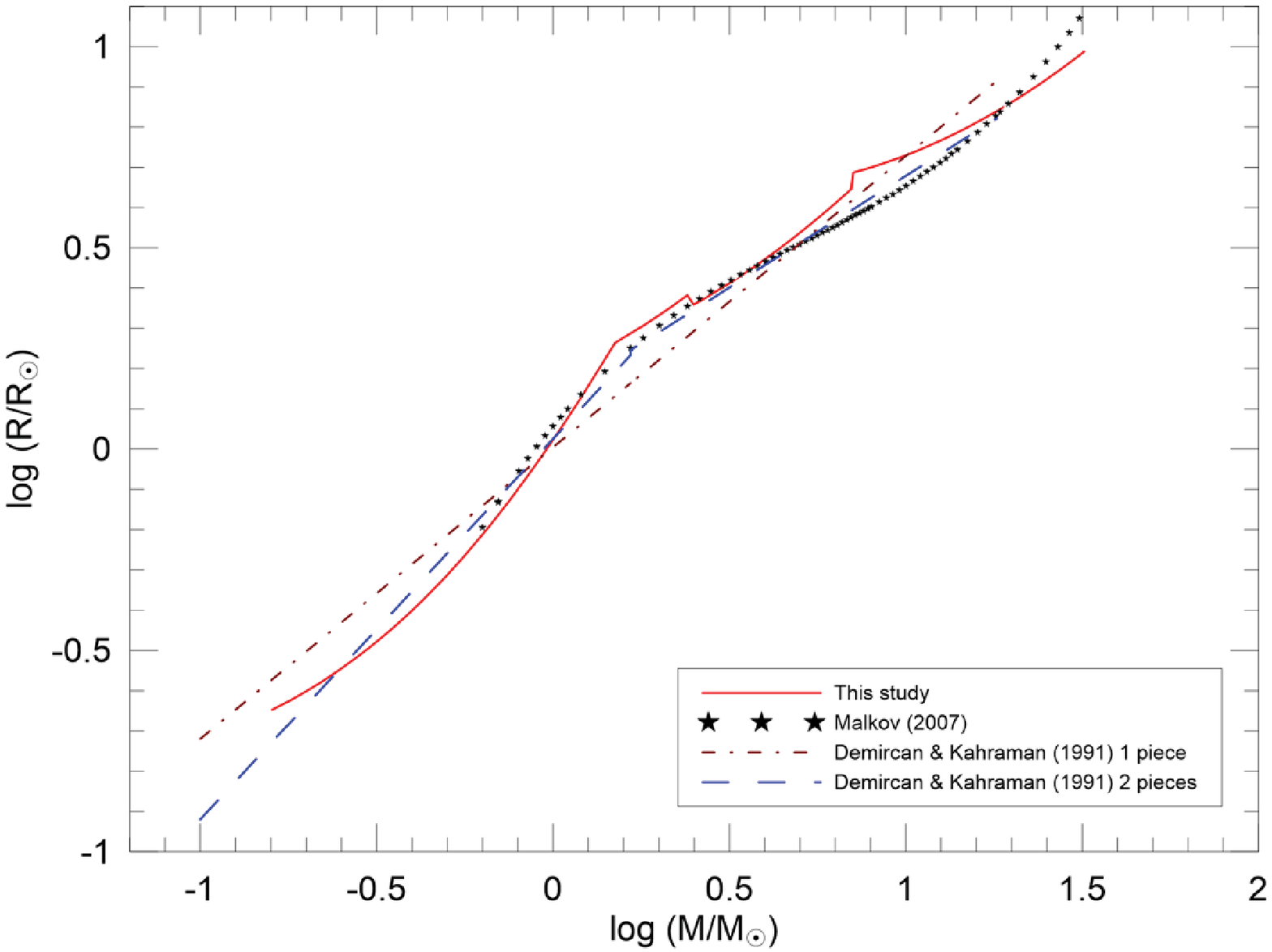}
\caption[] {Comparing calibrated MRR of this study to previous determinations.}
\end{center}
\end{figure*}

The calibrated MTR is compared to previous MTRs which are implied by the calibrated MLR and MRR of previous studies \citep{Malkov07, Demircan91} in Fig. 12. The MTR predicted from \citet{Malkov07}'s MLR and MRR shows better agreement to the MTR from this study, compared to the MTR calculated from \citet{Demircan91}'s MLR and MRR functions. It is very interesting to see temperature estimation of the two-piece models of \citet{Demircan91} for the stars $M<0.45M_{\odot}$, which over estimates the effective temperatures of low-mass stars. One may notice that the predicted temperatures are surprisingly parallel.

\begin{figure*}
\begin{center}
\includegraphics[scale=1, angle=0]{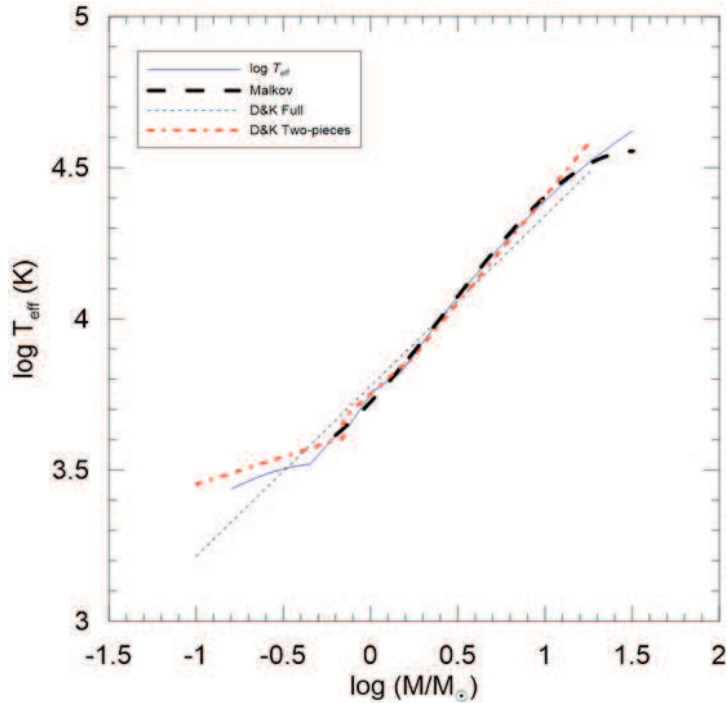}
\caption[] {Comparing calibrated MTR of this study to previous MTRs. Since previous MTRs were unavailable, they were computed from previous MLR and MRR functions.}
\end{center}
\end{figure*}

\subsection{Combining SOS survey and absolute parameters}

Sejong Open cluster Survey (SOS) is a dedicated project which provides homogeneous photometry from a large number of clusters in the SAAO Johnson-Cousins' $UBVI$ system by \citet{Sung13}. Galactic open clusters are stellar systems containing gravitationally bound stars, which may count numbering several tens to hundreds and having nearly the same chemical composition and age. Stars in small clusters and groups are dynamically unbound so the stars in small clusters and groups disperse and dispersed stars become the field stars. Studying photometry of the intermediate age open clusters is advantageous in that their mean colours and magnitudes can be used to represent the photometry of field stars, which were not included in this study. 

After observing a large number of open clusters in both the southern and northern hemispheres, \citet{Sung13} published a table giving typical spectral types (52 distinct types from O2 to M5), effective temperatures, $B-V$ and $U-B$ colours and bolometric corrections, for the luminosity classes Iab, III and V. Absolute parameters of the stars were not provided. This is because, absolute parameters such as masses and radii, actually, are not expected to be determined in the photometric surveys such as SOS. Reliable absolute parameters such as $M$, $R$ and $T_{eff}$ come only from the simultaneous solutions of radial velocity and light curves of the detached double-lined eclipsing binaries \citep{Andersen91, Torres10, Eker14}. Since considerable numbers of stars in this study are still members of some open clusters and the rest are the field stars, we may as well combine magnitudes and colours of main-sequence stars from \citet{Sung13} with well known absolute parameters of main-sequence stars in this study, which can be predicted from interrelated MLR, MRR and MTR functions.

It is customary when combining two tables of stellar data, one containing photometric data and the other the absolute parameters, to connect both tables by a common single column showing the spectral types as given in both tables, as done by \citet{Cox00} and as discussed in the introduction. Spectral types however, are the least reliable parameter to be determined from the combined solutions of radial velocity and light curves of the detached double-lined eclipsing binaries. It is usually estimated from the effective temperatures and the colours of the components by looking at pre-determined tables listing spectral types and temperatures and colours. Therefore, for this study we chose to use the column with $T_{eff}$ as the connecting column, rather than the column with spectral types.

Table 7 is produced by combining the photometric data of \citet{Sung13}, which is from a large number of open clusters with different ages and typical absolute parameters  adjusted from the  MLR, MRR and MTR functions calibrated in this study. The columns of Table 7 indicate spectral types, logarithm of $T_{eff}$, $B-V$, $U-B$, and bolometric correction in the first five columns and in the same order given by \citet{Sung13}. The rest of the columns are visual absolute  magnitude ($M_V$), $T_{eff}$, and then absolute parameters $M$, $R$, $\log g$, $M/L$ in Solar units and finally $L/M$ in cgs units. Unlike Table 6 which is listing the locus points according to increasing masses, Table 7 is opposite in displaying indicated columns as decreasing masses.

\begin{table}
\setlength{\tabcolsep}{4pt}
\centering
\caption{Spectral types, colours and bolometric corrections and mean absolute parameters of the main-sequence stars according to SOS and MLR, MRR, MTR functions defined in this study.}
\begin{tabular}{cccccccccccc}
\hline
Spt & $\log T_{eff}$ & $B-V$   & $U-B$   & BC    & $M_V$    & $T_{eff}$ & $M$         & $R$         &  $\log g$  & $M/L$   & $L/M$ \\
    &      & (mag)   & (mag)   & (mag) & (mag)    & (K)       & ($M_{\odot}$) & ($R_{\odot}$) &  (cgs) &   ($\odot$) & (erg s$^{-1}$ gr$^{-1}$)  \\
\hline
O2 & 4.720 & -0.33 & -1.22 & -4.52 & -6.44 & 52483 & 63.980 &16.734 & 3.80 & 0.00003 & 57628 \\
O3 & 4.672 & -0.33 & -1.21 & -4.19 & -5.62 & 46990 & 44.260 &12.312 & 3.90 & 0.00007 & 28981 \\
O4 & 4.636 & -0.33 & -1.20 & -3.94 & -5.13 & 43251 & 34.809 &10.301 & 3.95 & 0.00010 & 18514 \\
O5 & 4.610 & -0.33 & -1.19 & -3.77 & -4.80 & 40738 & 29.669 & 9.236 & 3.98 & 0.00014 & 13743 \\
O6 & 4.583 & -0.33 & -1.18 & -3.58 & -4.50 & 38282 & 25.380 & 8.362 & 4.00 & 0.00019 & 10270 \\
O7 & 4.554 & -0.33 & -1.17 & -3.39 & -4.20 & 35810 & 21.660 & 7.616 & 4.01 & 0.00025 & 7642 \\
O8 & 4.531 & -0.32 & -1.15 & -3.23 & -3.99 & 33963 & 19.215 & 7.131 & 4.02 & 0.00032 & 6111 \\
O9 & 4.508 & -0.32 & -1.13 & -3.03 & -3.83 & 32211 & 17.123 & 6.721 & 4.02 & 0.00039 & 4929 \\
B0 & 4.470 & -0.31 & -1.08 & -2.84 & -3.45 & 29512 & 14.277 & 6.171 & 4.01 & 0.00055 & 3511 \\
B1 & 4.400 & -0.28 & -0.98 & -2.40 & -2.93 & 25119 & 10.459 & 5.454 & 3.98 & 0.00098 & 1965 \\
B2 & 4.325 & -0.24 & -0.87 & -2.02 & -2.35 & 21135 &  7.699 & 4.967 & 3.93 & 0.00174 & 1110 \\
B3 & 4.265 & -0.21 & -0.75 & -1.62 & -1.68 & 18408 &  6.123 & 3.989 & 4.02 & 0.00373 & 518 \\
B5 & 4.180 & -0.17 & -0.58 & -1.22 & -0.76 & 15136 &  4.516 & 3.214 & 4.08 & 0.00928 & 208 \\
B6 & 4.145 & -0.15 & -0.50 & -1.02 & -0.44 & 13964 &  4.007 & 2.974 & 4.09 & 0.01327 & 146 \\
B7 & 4.115 & -0.13 & -0.43 & -0.85 & -0.18 & 13032 &  3.625 & 2.797 & 4.10 & 0.01790 & 108 \\
B8 & 4.080 & -0.11 & -0.35 & -0.66 &  0.13 & 12023 &  3.234 & 2.617 & 4.11 & 0.02518 & 77 \\
B9 & 4.028 & -0.07 & -0.19 & -0.39 &  0.57 & 10666 &  2.743 & 2.394 & 4.12 & 0.04121 & 47 \\
A0 & 3.995 & -0.01 & -0.01 & -0.24 &  0.86 &  9886 &  2.478 & 2.274 & 4.12 & 0.05587 & 35 \\
A1 & 3.974 &  0.02 &  0.03 & -0.15 &  0.90 &  9419 &  2.325 & 2.362 & 4.06 & 0.05897 & 33 \\
A2 & 3.958 &  0.05 &  0.06 & -0.08 &  1.06 &  9078 &  2.216 & 2.292 & 4.06 & 0.06919 & 28 \\
A3 & 3.942 &  0.08 &  0.08 & -0.03 &  1.23 &  8750 &  2.113 & 2.226 & 4.07 & 0.08107 & 24 \\
A5 & 3.915 &  0.15 &  0.10 &  0.00 &  1.57 &  8222 &  1.952 & 2.123 & 4.08 & 0.10557 & 18 \\
A6 & 3.902 &  0.18 &  0.10 &  0.01 &  1.74 &  7980 &  1.879 & 2.077 & 4.08 & 0.11971 & 16 \\
A7 & 3.889 &  0.21 &  0.09 &  0.02 &  1.91 &  7745 &  1.810 & 2.033 & 4.08 & 0.13563 & 14 \\
A8 & 3.877 &  0.25 &  0.08 &  0.02 &  2.07 &  7534 &  1.749 & 1.994 & 4.08 & 0.15209 & 13 \\
F0 & 3.855 &  0.31 &  0.05 &  0.01 &  2.37 &  7161 &  1.643 & 1.928 & 4.08 & 0.18726 & 10 \\
F1 & 3.843 &  0.34 &  0.02 &  0.01 &  2.53 &  6966 &  1.588 & 1.893 & 4.08 & 0.20956 & 9.22 \\
F2 & 3.832 &  0.37 &  0.00 &  0.00 &  2.69 &  6792 &  1.540 & 1.863 & 4.09 & 0.23219 & 8.33 \\
F3 & 3.822 &  0.40 & -0.01 &  0.00 &  2.82 &  6637 &  1.498 & 1.838 & 4.09 & 0.25448 & 7.60 \\
F5 & 3.806 &  0.45 & -0.02 & -0.01 &  3.30 &  6397 &  1.354 & 1.588 & 4.17 & 0.35692 & 5.42 \\
F6 & 3.800 &  0.48 & -0.01 & -0.02 &  3.49 &  6310 &  1.305 & 1.508 & 4.20 & 0.40325 & 4.79 \\
F7 & 3.794 &  0.50 &  0.00 & -0.02 &  3.65 &  6223 &  1.259 & 1.434 & 4.23 & 0.45450 & 4.25 \\
F8 & 3.789 &  0.53 &  0.02 & -0.03 &  3.80 &  6152 &  1.222 & 1.377 & 4.25 & 0.50125 & 3.86 \\
G0 & 3.780 &  0.59 &  0.07 & -0.04 &  4.06 &  6026 &  1.161 & 1.283 & 4.29 & 0.59553 & 3.25 \\
G1 & 3.775 &  0.61 &  0.09 & -0.04 &  4.19 &  5957 &  1.128 & 1.236 & 4.31 & 0.65401 & 2.96 \\
G2 & 3.770 &  0.63 &  0.13 & -0.05 &  4.33 &  5888 &  1.098 & 1.191 & 4.33 & 0.71723 & 2.70 \\
G3 & 3.767 &  0.65 &  0.15 & -0.06 &  4.42 &  5848 &  1.080 & 1.165 & 4.34 & 0.75756 & 2.55 \\
G5 & 3.759 &  0.68 &  0.21 & -0.07 &  4.64 &  5741 &  1.031 & 1.097 & 4.37 & 0.87824 & 2.20 \\
G6 & 3.755 &  0.70 &  0.23 & -0.08 &  4.72 &  5689 &  1.019 & 1.081 & 4.38 & 0.92834 & 2.08 \\
G7 & 3.752 &  0.72 &  0.26 & -0.09 &  4.78 &  5649 &  1.011 & 1.069 & 4.39 & 0.96765 & 2.00 \\
G8 & 3.745 &  0.74 &  0.30 & -0.10 &  4.92 &  5559 &  0.990 & 1.041 & 4.40 & 1.06553 & 1.81 \\
K0 & 3.720 &  0.81 &  0.45 & -0.18 &  5.45 &  5248 &  0.922 & 0.951 & 4.45 & 1.49639 & 1.29 \\
K1 & 3.705 &  0.86 &  0.54 & -0.24 &  5.77 &  5070 &  0.884 & 0.903 & 4.47 & 1.82840 & 1.06 \\
K2 & 3.690 &  0.91 &  0.65 & -0.32 &  6.11 &  4898 &  0.848 & 0.858 & 4.50 & 2.22864 & 0.867 \\
K3 & 3.675 &  0.96 &  0.77 & -0.41 &  6.46 &  4732 &  0.813 & 0.817 & 4.52 & 2.71007 & 0.713 \\
K5 & 3.638 &  1.15 &  1.06 & -0.65 &  7.32 &  4345 &  0.736 & 0.727 & 4.58 & 4.34779 & 0.445 \\
M0 & 3.580 &  1.40 &  1.23 & -1.18 &  9.07 &  3802 &  0.558 & 0.541 & 4.72 & 10.16323 & 0.190 \\
M1 & 3.562 &  1.47 &  1.21 & -1.39 &  9.60 &  3648 &  0.524 & 0.508 & 4.75 & 12.75337 & 0.152 \\
M2 & 3.544 &  1.49 &  1.18 & -1.64 & 10.15 &  3499 &  0.492 & 0.479 & 4.77 & 15.91749 & 0.121 \\
M3 & 3.525 &  1.53 &  1.15 & -2.02 & 10.85 &  3350 &  0.462 & 0.452 & 4.79 & 20.00404 & 0.097 \\
M4 & 3.498 &  1.56 &  1.14 & -2.55 & 12.29 &  3148 &  0.323 & 0.338 & 4.89 & 32.12216 & 0.060 \\
M5 & 3.477 &  1.61 &  1.19 & -3.05 & 13.37 &  2999 &  0.249 & 0.284 & 4.93 & 42.55068 & 0.045 \\
\hline
\end{tabular}%
\end{table}%

For the stars $M>1.5 M_{\odot}$, $T_{eff}(M)$ function in Table 5 was used to calculate $T_{eff}$ in column 7. By trial and error, that is, different values of $M$ was tried until proper $T_{eff}$ is produced which is given in the second column kept logarithmic as its original form in \citet{Sung13}. Once a typical $M$ is produced in accord with the spectral types and $\log T_{eff}$ (first and second columns), then typical $L$ (not listed) is calculated from the MLR function of this study. Consequently, the typical $L$ is converted to $M_V$ using the bolometric correction given in column 5. $M_V$, then, is listed in column 6. Morever, using typical $L$ (not listed) and $T_{eff}$ in column 7, typical $R$ is calculated according to $L=4\pi R^2 \sigma T^4_{eff}$. Finally, the last three columns are produced from the typical luminosities and masses.  

For the stars $M\leq 1.5M_{\odot}$, method of computing absolute parameters changes because there is no calibrated $T_{eff}(M)$ function for the low mass stars. In order to compute typical $T_{eff}$ for low mass stars $M\leq 1.5M_{\odot}$, both the $R(M)$ function given in Table 5 and the $L(M)$ functions given in Table 4 are used. Again,  various values of $M$ were tried until $T_{eff}$ which has same $\log T_{eff}$ value in the same row were found from the relation $L=4\pi R^2 \sigma T^4_{eff}$, where $L$ and $R$ comes from $R(M)$ and $L(M)$. 

For the most massive O2, O3, and O4 stars shown in the first three rows of Table 7, the calculated masses of 64, 44, and $35M_{\odot}$ all exceed the $31M_{\odot}$ upper limit of the mass range studied here. Note however, that those masses are extrapolated values, based on the $T_{eff}(M)$ and $L(M)$ functions and corresponding radii. Rather than leaving the first three rows of Table 7 empty, we chose to show the extrapolated values. Readers interested in the most massive stars in particular should be aware of the difference.

\subsection{More accurate data or more data?}

At earlier times, when observational data were limited, astronomers were collecting all data without paying attention to its quality. When determining their MRR, \citet{Gimenez85} used observed radii from five resolved binaries, 14 visual binaries and 12 OB binaries with less accuracy. \citet{Demircan91} used observational data of 70 eclipsing binaries including the ones with main-sequence components of detached, and semi-detached and OB-type contact and near-contact systems. Especially after critical compilation of absolute dimensions of binary components by \citet{Popper67, Popper80}, \citet{Andersen91} was the first author, who was very selective when collecting detached double-lined eclipsing systems having masses and radii within 2\% uncertainty. \citet{Gorda98} too collected stellar masses and radii with accuracies within 2-3\% from photometric, geometric, and absolute elements of 112 eclipsing binaries with both components on the main sequence. While \citet{Henry93} and \citet{Malkov07} were a little bit more tolerant accepting accuracies 15\% and 10\% respectively, \citet{Torres10} collected masses and radii of eclipsing binaries within 3\% in order to study MLR and MRR diagrams. In our previous study \citep{Eker15}, we have followed the trend of preferring the highest accuracy when collecting absolute parameters of detached eclipsing double-lined spectroscopic binaries with both $M$ and $R$ accuracy $\leq 3\%$ and accuracy of $L\leq30\%$ when calibrating a four piece MLR. 

In this study, however, we have decided not to follow the same trend because we have noticed that the accuracy of the effective temperature of a star, if computed from its $M$ and $R$ using a calibrated MLR, depends mostly on the luminosity dispersions rather than the observational random errors of $M$ and $R$. If a bolometric luminosity is computed from $R$ and $T_{eff}$, through the Stefan-Boltzmann law, the propagated random error of the luminosity is much larger than the observational random errors of $M$ and $R$ since

\begin{equation}
{\frac{\Delta L}{L}}=2{\frac{\Delta R}{R}}+4{\frac{\Delta T}{T}}.
\end{equation}
It makes no difference whether or not measurements of $L$ come from the observed $R$ and $T_{eff}$ as in the case of eclipsing binaries or directly from star's bolometric absolute magnitude if its parallax was known. In the latter case, huge parallax errors and error in bolometric correction may make $\Delta L/L$ even larger. The position of a main-sequence star on a $\log M - \log L$ diagram does not only depend on observational parameters but also depends on its chemical composition and evolution (age). In fact, distribution is more affected by  chemical compositions and ages than random observational errors. Thus, if $L$ is predicted from a MLR function according to $M$, then uncertainty contributions of observational parameters are negligible when compared to the uncertainty contributions from the chemical composition and evolution \citep{Andersen91, Torres10, Eker15}. Using more accurate $M$ does not improve the predicted value of $L$. On the contrary, there is a tolerance limit for $M$, which could be calculated as

\begin{equation}
{\frac{\Delta M}{M}}={\frac{1}{\alpha}}{\frac{\Delta L}{L}}
\end{equation}
where $\Delta L/L = \sigma/0.4343$, in which $\sigma$ is the standard deviation of data on the $\log M - \log L$ diagram. Unless observational uncertainty of $M$ is greater than the tolerance, the predicted $L$ and its relative uncertainty would be the same for a given MLR function. For example, let us assume the standard deviation in very high-mass domain is $\sigma=0.158$, as given in Table 3 of \citet{Eker15}. The relative uncertainty of $L$ due to this dispersion would be 36\% according to $\Delta L/L=\sigma /0.4343$. If the power of $M$ ($L \propto M^{\alpha}$) is $\alpha= 2.726$ as also listed in the same table of \citet{Eker15}, the tolerance of $M$ is about 13\%. This means that, unless observational error of $M$ is more than 13\%, the relative uncertainty of $L$ is the same, because it is determined by existing dispersion ($\Delta L/L = \sigma/0.4343$) even if the value of $M$ is errorless. 

Therefore, limiting $M$ and $R$ accuracies to a smaller percentage (such as 3\%) will cause loss of data rather than a gain in information. According to Table 5 of \citet{Eker15}, the minimum tolerance is about 6\% for stars in the low mass and intermediate-mass domains. Therefore, for this study, we chose to set our limiting accuracy to at least 6\% when selecting the 509 main-sequence stars to use for determining and calibrating the MLR, MRR and MTR functions obtained.    

Obviously the number of data is very important because each single datum contributes according to its accuracy, precision and/or position on the diagrams. Eliminating some data according to a limiting accuracy may discard crucial information, so that studying with the highest accuracy may not always mean gaining more information. So we decided not to discard stars with $M$ and $R$ accuracies up to 15\% in order not to lose information. Lowering the limiting accuracy from 3\% to 6\% caused us to gain 88 stars. Setting the limit to 15\% made us gain 76 more stars. Otherwise, with a 3\% limit, the present sample would have numbered only 345 stars, compared to the 509 stars in the present sample. 

Adding more stars permitted us to define the ultra low-mass domain ($0.179\leq M/M_{\odot}\leq 0.45$). According to Table 4, there are 22 stars in the ultra low-mass domain. Among those 22, the eight stars have $M$ and $R$ accuracies $\leq 3\%$; 10 stars have $M$ and $R$ accuracies 3-6\% and four stars have $M$ and $R$ accuracies 6-15\%. That is, if we did not include lower quality data, the number of stars in the ultra low-mass domain would have been eight, which were already included in the list of \citet{Eker15}. Five of them appeared as if in the low mass domain, and three of them appeared not obeying the MLR of other low-mass stars. Only after adding the additional 14 stars with less accurate $M$ and $R$ in this study, are we able to define the new domain for main-sequence stars of ultra low-mass. 

\subsection{Uniqueness of MLR, MRR and MTR} 
Some authors \citep{Andersen91, Henry93, Torres10} preferred not to define MLR because the scatter on the mass-luminosity diagram is not due to observational errors but most likely abundance and evolutionary effects. \citet[][p.107]{Andersen91}, claims ``... departures from a unique relation are real''. If there is no unique function to represent MLR of main-sequence stars, why bother to define one?

Obviously, clarification of the uniqueness problem attributed to  MLR is necessary. \citet{Eker99}, who analyzed the uniqueness problem of star spot models, claimed that a scientist, who evaluates any scientific data, may face three types of uniqueness problems (type 1, type 2 and type 3). Type 1 is the non-uniqueness of the function to generate a curve to fit the data. Type 2 is the non-uniqueness of the fit, and type 3 is the non-uniqueness of the parameter space. All three problems are sequentially inter-connected. To be free of a uniqueness problem, the answer ``yes'' is required for the following questions: 1) is the function unique? 2) Is a unique fit possible? 3) Is the parameter space unique? If any one of the answers is ``no'', a uniqueness problem is inevitable.

Confusion between different meanings of the word ``unique'' result in ambiguity. Non-uniqueness of the function according to the first question is not the non-uniqueness implied by \citet{Andersen91} and \citet{Torres10}. Choosing a function in most physical problems does not pose a problem because there are certain mathematical expressions for certain data, e.g. the Planck function to express the spectral energy distribution of stars. Only if the data is not known to be associated with a specific function, and if scientists are confused between the two or more functions to generate a fitting curve, can non-uniqueness of type 1 be claimed. First of all, the basic function to express a MLR is a power law ($L \propto M^{\alpha}$). However, different functions suggested by different authors to fit the data on $\log M - \log L$ plane do not pose a problem because different functions imply how a real number ($\alpha$), the power of $M$, changes at different domains of $M$. Therefore, non-uniqueness of type 1 does not exist. The second question is answered ``yes'' because there are methods, like the least squares method, which guarantees the uniqueness of the fit. The third question is answered ``yes'' because of regardless the value of $\alpha$, there so is only one $L$, for a given $M$. Thus, the parameter space is also unique. The non-uniqueness implied by \citet{Andersen91} and \citet{Torres10} does not apply to MLR relations.

When a main-sequence mass-luminosity relation was suggested first in 1923 \citep{Hertzsprung23, Russell23}, the theory of stellar structure and evolution was not yet fully established. Astrophysicists had to wait until 1932, for the discovery of the neutron \citep{Chadwick33} in order to establish nuclear fusion as a source for stellar energy. Only after the CNO cycle reactions were established by Hans Bethe and Von Weizsacher \citep{Clayton68}, and only a year later the p-p chain reactions were established by \citet{Bethe39}, were the solutions of stellar structure equations together with the nuclear energy equations able to place our theoretical understanding of the evolution of stars on solid ground \citep{Clayton68}. That is, the observational discovery MLR in the middle of the first half of the 20th century was confirmed later theoretically that mass ($M$) is the prime parameter which determines internal structure, size ($R$) and luminosity ($L$) of a star not only for the time span of main-sequence but also throughout star's lifetime until its death, where initial chemical composition can cause little variations. So, the scatter on a mass-luminosity diagram for field main-sequence stars is not only due to observational errors, but also due to various ages and chemical compositions.

Unless it is established for main-sequence stars in a Galactic open cluster, where all stars have the same age and metallicity, according to \citet{Andersen91} and \citet{Torres10}, there could be a countless number of MLRs, because each combination of metallicity and age implies a different MLR. This way of thinking, however, leads scientists to the idea that MLRs are undefinable and/or useless. 

At this point, we could still argue that the heterogeneous nature of the data do not change the general characteristics of a typical MLR function (or a MRR or MTR), so that an statistically determined function for a given sample, e.g. in the Solar neighborhood stars which are known to be main-sequence stars mostly Solar metallicity distributed within $0.008<Z<0.040$, must be unique and useful for many practical applications. Stellar evolution theory does not discredit MLR determinations, but tells us that the single value of $L$ for a given $M$ is a unique value, a kind of a mean $L$ of different metallicities and ages existing in the sample of a given $M$. Since there is only one value of $L$ for a given $M$, there must be degeneracy in $L$. Parameters to break up this degeneracy are metallicity and age. In other words, it should be possible to obtain the true $L$ of any main-sequence star from any sample using a proper evolutionary line (track) if the star's age and chemical composition are known.

Please note that if there is no age and/or metallicity measurement for a star in the sample, it is not a problem of uniqueness; rather it is a problem of degeneracy for that individual star. The mass-brightness relations of a given age and metallicity, which could be stated for an open or globular cluster, are called isochrones. Isochrones, however, cannot be considered as main-sequence MLR, according to the concept described in this study which was originally introduced by \citet{Hertzsprung23} and \citet{Russell23} in the first half of the 20th century.      

\section{Conclusions}

An update of the Catalogue of Stellar Parameters from the Detached Double lined eclipsing binaries \citep{Eker14} has been provided. It includes 64 new binaries, and one new SB3 triple system, and increases the number of stars in the catalogue from 514 (257 binaries) to 639 (318 binaries plus one triple). In addition to increasing the quantity of data in the catalogue, the quality of the data has also been improved. The number of stars having $M$ and $R$ measurements with better than 3\% accuracy is increased from 311 to 400, and with better than 5\% accuracy is increased from 388 to 480. From the 639 stars now available in the updated catalogue, 509 main-sequence stars, with $M$, $R$ and $T_{eff}$ values accurate to within 15\% have been selected for the present study.

Interrelated MLR, MRR and MTR relations have been calibrated using the published parameters ($M$, $R$ and $T_{eff}$) for all 509 main-sequence stars selected. A six-piece classical MLR ($L \propto M^{\alpha}$) is calibrated based on the $\log M - \log L$ diagram in the full range of masses ($0.179\leq M/M_{\odot}\leq 31$). A quadratic MRR is  calibrated on a diagram of linear scale using published radii directly in the mass range $0.179\leq M/M_{\odot}\leq 1.5$. The MTR is calibrated on the logarithmic scale using published $T_{eff}$ in the mass range $1.5\leq M/M_{\odot}\leq 31$. Missing parts of the MRR were completed using the Stefan-Boltzmann law ($L=4\pi R^2 \sigma T^4$) and calibrated MLR and MTR, while the missing parts of the MTR were completed using the Stefan-Boltzmann law ($L=4\pi R^2 \sigma T^4_{eff}$) and calibrated MLR and MRR, so interrelated MLR, MRR and MTR were obtained. 

The interrelated MLR, MRR and MTR functions and the present data allowed us to serendipitously discover new break points on the $\log M - \log L$ diagram. A total of five break points separating the main-sequence mass domains an in which the classical MLR has a constant power ($L \propto M^{\alpha}$), were identified and discussed. F-test results applied to the $M-L$ diagram indicate that a six-piece linear MLR function is equivalent to a fifth or a sixth degree polynomial. Because the coefficients of such polynomials are physically meaningless, and using such a polynomial is relatively impractical, the six-piece linear MLR is found not only physically more meaningful but also more practical. The interrelated MLR, MRR, MTR and break points were confirmed not only by observational data but also by the locus points of main-sequence stars in the Solar vicinity and Galactic disc. 

The interrelated MLR, MRR and MTR functions obtained in this study can now be used to determine the absolute properties of main-sequence stars in other samples and in general. Based purely on the observations of another sample of main-sequence stars published by \citet{Sung13} for example, including colours ($B-V$, $U-B$) and magnitudes, as well as $T_{eff}$ and bolometric corrections for spectral type, the absolute properties were determined based on the new MLR, MRR and MTR functions, including stellar $M$, $R$, $T_{eff}$, $\log g$, $M/L$ and $L/M$. The column of $T_{eff}$ was used as the connecting column when combining Table 5 of \citet{Sung13} containing photometric data with new the columns showing absolute data determined in this study.  

A key feature of the present study, compared to previous ones including our own \citep{Eker15}, is its inclusion of more stars with $M$ and $R$ measurements of lower accuracy, particular in the lowest mass domain. Whereas previous studies have typically considered only stars with $M$ and $R$ known to 3\% accuracy, as in \citet{Eker15}, the inclusion in this study of stars with $M$ and $R$ of 15\% accuracy has allowed us to significantly extend the interrelated MLR, MRR and MTR functions from a lower mass limit of $0.38M_{\odot}$ to $0.179M_{\odot}$. 

The uniqueness problem of scientific activities, which may also be applied to the interrelated MLR, MRR and MTR functions in this study, were discussed. On analysis, it can be concluded that there is not any kind of uniqueness problem with the newly determined MLR, MRR and MTR. These functions must be unique in order to represent the present sample of main-sequence stars.   

\section{Acknowledgments}
This work has been supported in part by the Scientific and Technological Research Council (T\"UB\.ITAK) by the grant number 114R072. Thanks to Akdeniz University BAP office for providing a partial support for this research. Olcay Plevne contributed in computing F-test tables. The authors would also like to thank the anonymous referee for providing valuable comments that improved the quality and presentation of the paper.

\end{document}